\documentclass[aps,12pt]{revtex4-2}

\usepackage{graphicx}
\usepackage{dcolumn}
\usepackage{amsmath}
\usepackage{amssymb}
\usepackage{amsbsy}
\usepackage{float}
\usepackage{url}
\usepackage[bookmarksopen]{hyperref}
\usepackage{caption}
\usepackage{physics}
\usepackage{float}
\usepackage{natbib}

\begin{document}

\title{Wave Is All You Need: A Quantum Language Model Based on Fully Connected Quantum Boltzmann Machine}

\author{Liu Jun}
\email{liujun@hufe.edu.cn}
\email{lewjhon@126.com}
\affiliation{School of Economics, Hunan University of Finance and Economics, Changsha 410205, China}

\begin{abstract}To address the limitations of the classical self-attention mechanism,
this work proposes a self-attention mechanism based on the fully connected quantum Boltzmann machine (FC-QBM), and further constructs a fully connected quantum Boltzmann machine language model (FC-QBM-LM) using the mechanism as the core engine. By performing unitary evolution on semantic quantum states, the mechanism implicitly implements positional encoding and self-attention computation, eliminating the need to decompose continuous semantic streams into discrete matrices for explicit matrix multiplication, with the measurement results naturally satisfying the normalization condition. Specifically, this work rigorously derives the computational principle of the self-attention mechanism based on FC-QBM. It demonstrates that this mechanism, in which FC-QBM encodes token positions and interactions and \(R_\mathrm{Y}\) gates perform differentiated information extraction, is mathematically equivalent to the classical self-attention mechanism. The mathematical principles governing the training and generation of FC-QBM-LM are thoroughly derived, and the two corresponding algorithms are subsequently developed. An 8-qubit FC-QBM-LM is finally constructed and trained on the corpus of \textit{Three Hundred Tang Poems}, followed by computational experiments using the model. Under the 8-qubit setting, the quantum layer implements the entire positional encoding and self-attention computation pipeline with only 44 parameters. In the noiseless scenario, the model exhibits favorable convergence. Furthermore, the quantum circuit proposed in this work, which terminates with \(R_{\mathrm{Y}}\) gates, achieves better convergence than the counterpart terminating with \(R_{\mathrm{X}}\) gates. Under the small-sample regime and across eight distinct noise intensity levels, the model demonstrates strong noise robustness, retaining over 99\% of its noiseless convergence performance. Experiments on five-character quatrain generation reveal that the model preliminarily exhibits three core capabilities: associative and divergent thinking, content generation across imagery and human-aligned word ordering, validating the effectiveness of the proposed self-attention mechanism based on FC-QBM.
    
\end{abstract}

	\maketitle
	\section{Introduction}

The evolution of language models is essentially a process of continuous advancement in semantic modeling capabilities. Early rule-based language models relied on manually constructed formal grammars and dictionaries\cite{chomsky1957syntactic}. Constrained by scalability bottlenecks, they struggled to characterize the complex properties of real-world language. In the 1990s, statistical language models exemplified by \(n\)‑gram and hidden Markov models (HMMs) enabled probabilistic prediction of word sequences using large-scale corpora and greatly improved the performance of tasks such as machine translation\cite{brants2007large}. Nevertheless, they were plagued by the curse of dimensionality. At the beginning of the 21st century, neural network language models pioneered vector representations. Bengio et al. (2003) first mapped words into a low-dimensional continuous vector space\cite{bengio2003neural}. Recurrent neural networks (RNNs) and long short-term memory networks (LSTMs) captured long-range dependencies through gating mechanisms\cite{hochreiter1997long}. However, sequential computation hindered parallel processing, and long-term memory decay also limited model scaling. In 2017, the transformer architecture adopted the self-attention mechanism to implement parallel modeling of global sequence dependencies\cite{vaswani2017attention}, thoroughly breaking the performance bottleneck of recurrent architectures and directly promoting the explosive development of large language models (LLMs).

However, the classical self-attention mechanism suffers from two fundamental limitations. First, within the conventional framework, the computation of semantic interactions between tokens can only be realized through large-scale matrix operations. Such matrix computation based on discrete token sets inherently undermines the continuity and integrity of semantic representations. A study indicates that the underlying semantics of natural language essentially lie on a continuous manifold, and discretization leads to the loss of manifold smoothness\cite{mabrok2026latent}. Second, the implementation approach of the classical self-attention mechanism, which relies on massive matrix operations, further introduces substantial computational overhead. The mechanism requires repeated execution of numerous matrix multiplications, and its time and space complexity grows quadratically with the sequence length. Notably, even approximate computation, as a common strategy to reduce such quadratic overhead, faces theoretical constraints. Under the strong exponential time hypothesis, it is impossible to construct subquadratic approximate self-attention algorithms with rigorous error guarantees\cite{keles2023computational}. Given the current technical conditions, training and inference typically rely on large-scale or even ultra-large-scale GPU clusters, resulting in prohibitive computational resource consumption.

To address the aforementioned limitations of the classical self-attention mechanism, continuous semantic modeling has emerged as an important breakthrough direction. Diffusion-LM pioneered research on diffusion language modeling in the embedding space\cite{li2022diffusion}, while Difformer proposed an improved scheme addressing critical drawbacks such as training instability exhibited by Diffusion-LM\cite{gao2024empowering}. Hu et al. (2026) proposed embedded language flows (ELF), which constructs diffusion language models within the continuous embedding space and maps continuous vectors to discrete tokens only at the final generation step. This design preserves the continuity of semantic representations and enables direct transfer of well-established diffusion techniques from the vision domain, achieving substantial improvements in generation quality and sampling efficiency over discrete diffusion language models\cite{hu2026elf}. While these attempts mitigate the loss of semantic continuity induced by matrix operations based on discrete token sets, they remain largely rooted in the transformer architecture and fail to substantially address the computational complexity bottleneck.

Quantum computing, leveraging superposition and entanglement, opens up a new avenue for performance breakthroughs in language models. Research on quantum language models has grown rapidly in recent years, yet most mainstream works remain limited to peripheral enhancement and partial substitution within the classical transformer framework. A survey has systematically summarized the acceleration potential of quantum computing in the training, inference and optimization of machine learning models\cite{biamonte2017quantum}. Hybrid quantum transformers represented by HyQuT integrate variational quantum circuits into the transformer framework to achieve local parameter compression\cite{kong2025hybrid}. Nevertheless, their self-attention computation still follows classical schemes and does not redefine the matrix operation logic for token interactions. Another review in quantum natural language processing also notes that most existing quantum language models fail to innovate upon the underlying computational framework of classical LLMs\cite{nausheen2025quantum}. Overall, existing quantum language models mostly leverage quantum computing to assist and augment the original classical processing pipeline, leaving the core architecture and computational logic of the self-attention mechanism fundamentally unaltered, thus rendering it infeasible to resolve the two inherent limitations of classical self-attention at their root.

By contrast, in the fields of probabilistic graphical models and energy-based learning (EBL), Boltzmann machines (BMs) offer an alternative route for global dependency modeling distinct from matrix dot-product operations. The foundational work on BMs originates from Hinton and colleagues\cite{hinton1983optimal,ackley1985learning}. The original fully connected Boltzmann machine (FC-BM) proposed in 1985 can theoretically fit any finite discrete joint distribution via bidirectional coupling among units\cite{ackley1985learning}. Nevertheless, computing the partition function and performing Gibbs sampling incur overheads scaling exponentially with the number of units, making FC-BMs impractical for deployment on classical computing hardware. To mitigate this bottleneck, Smolensky (1986) proposed the restricted Boltzmann machine (RBM) by introducing a bipartite graph topology without intra-layer connections, substantially reducing the model’s computational complexity\cite{smolensky1986information}. Building upon this, Hinton et al. (2006) constructed deep belief networks (DBNs) by stacking RBMs layer-wise, fueling the resurgence of deep learning\cite{hinton2006fast}. However, while simplifying the structure to enable feasible training, RBMs abandon the inherent pairwise coupling among units in fully connected architectures and partially lose the capacity to directly characterize high-order variable correlations. Recently, Kim and Park (2026) proposed Boltzmann attention, which introduces learnable Ising couplings into the transformer framework\cite{kim2026boltzmann}. They conducted experiments with exact Boltzmann computation on small-scale tasks, yet the exponential cost still prevents scaling toward large sequences. It follows that although FC-BMs possess prominent theoretical potential for interaction modeling, they have long been constrained by the inherent limitations of classical computing.

Quantum Boltzmann machines (QBMs) offer a viable route to tackle the aforementioned dilemma. Wiebe et al. (2014) demonstrated that quantum approaches can not only accelerate the training of deep RBMs but also enable efficient training of multi-layer FC-BMs\cite{wiebe2014quantum}. Benedetti et al. (2017) further illustrated the feasibility of embedding probabilistic graphical models into quantum-assisted learning hardware\cite{benedetti2017quantum}. Amin et al. (2018) replaced classical binary units with qubits and leveraged quantum annealing to achieve efficient sampling from the Boltzmann distribution\cite{amin2018quantum}. They also explored the possibility of verifying the model’s generative modeling capability on quantum annealing hardware. Liu (2026) proposed a bilevel optimization training scheme based on the quantum approximate optimization algorithm (QAOA)\cite{liu2026breaking}. Breaking the constraint of the fixed target Hamiltonian in standard QAOA, the scheme simulates the contrastive divergence learning through inner-loop and outer-loop computational procedures, enabling the training of fully connected quantum Boltzmann machines (FC-QBMs) using only a single layer of QAOA while delivering favorable noise robustness in noisy simulation settings. These studies indicate that quantum computing holds the potential to resolve the training bottleneck of FC-BMs. Nevertheless, to date, no work has deeply integrated FC-QBMs with the self-attention mechanism to reconstruct the computational architecture of attention.

Based on the above discussions, this paper attempts to reconstruct the classical self-attention mechanism at the fundamental level to overcome its inherent limitations. Specifically, a novel self-attention mechanism based on FC-QBM is designed, and a quantum language model is further constructed using the proposed mechanism as its core engine. The organization of this paper is as follows. First, on the basis of a brief review of the classical self-attention mechanism, the computational principle of the FC-QBM self-attention mechanism is derived, and its functional equivalence to the classical counterpart is mathematically proven. Second, the construction and implementation processes of the quantum language model based on FC-QBM are elaborated; the mathematical principles underlying model training and text generation are derived; and the corresponding algorithms are developed. Third, a fully connected quantum Boltzmann machine language model (FC-QBM-LM) with eight qubits is constructed based on the proposed algorithms. Finally, the model is trained on the corpus of \textit{Three Hundred Tang Poems} via the PennyLane quantum simulation platform and the PyTorch machine learning framework, followed by a comprehensive evaluation of its three core properties: convergence performance, noise robustness and five-character quatrain generation capability.

\raggedbottom	

\section{Fully Connected Quantum Boltzmann Machine Self-Attention Mechanism and Its Equivalence to Classical Self-Attention Mechanism}
\subsection{Review of Classical Self-Attention Mechanism}

The self-attention mechanism serves as the core module for transformer-based language models to perform contextual semantic modeling. It endows the model with the ability to capture long-range textual dependencies and perceive word order, imagery and grammatical correlations, enabling the model to understand and generate natural language. Based on the classical transformer architecture proposed by Vaswani et al. (2017), this paper first briefly reviews the basic principle and computational procedures of the self-attention mechanism\cite{vaswani2017attention}. Define a token sequence \(x_0,x_1,\dots,x_{n-1}\) of length \(n\), where each token is denoted as \(x_t\ (t=0,1,\dots,n-1)\). Let \(\boldsymbol{E}\in\mathbb{R}^{V\times d_{\text{model}}}\) denote the embedding matrix, where \(V\) is the total vocabulary size and \(d_{\text{model}}\) denotes the dimension of embeddings. Each row of \(\boldsymbol{E}\) is a \(d_{\text{model}}\)-dimensional vector, denoted as \(\boldsymbol{e}(x_t)\). Scaling \(\boldsymbol{e}(x_t)\) by the factor \(\sqrt{d_{\text{model}}}\) produces the embedding vector \(\boldsymbol{emb}(x_t)\) of token \(x_t\):
\begin{equation}
\label{eq1}
\boldsymbol{emb}(x_t) = \sqrt{d_{\mathrm{model}}} \cdot \boldsymbol{e}(x_t), \quad t \in \{0,1,\dots,n-1\}
\end{equation}

Stacking the embedding vectors of all tokens row by row yields the embedding matrix \(\boldsymbol{X}^{\text{emb}}\) for the entire token sequence:
\begin{equation}
\label{eq2}
\boldsymbol{X}^{\mathrm{emb}}
=
\begin{bmatrix}
\boldsymbol{emb}(x_0) \\
\boldsymbol{emb}(x_1) \\
\vdots \\
\boldsymbol{emb}(x_{n-1})
\end{bmatrix},\quad
\boldsymbol{X}^{\mathrm{emb}} \in \mathbb{R}^{n\times d_{\mathrm{model}}}
\end{equation}

The positional encoding is subsequently applied, with the specific rule from the original paper as follows:\hfill\mbox{}
\begin{equation}
\label{eq3}
\begin{aligned}
PE_{(pos,2i)} &= \sin\left( \frac{pos}{10000^{\left(2i/d_{\mathrm{model}}\right)}} \right) \\
PE_{(pos,2i+1)} &= \cos\left( \frac{pos}{10000^{\left(2i/d_{\mathrm{model}}\right)}} \right)
\end{aligned}
\end{equation}

where \(pos\) denotes the position of the token in the sequence, taking values of \(0,1,\dots,n-1\); \(i\) denotes the dimension index of embeddings, taking values of \(0,1,\dots,d_{\text{model}}/2-1\). The sequence’s positional encoding matrix \(\boldsymbol{PE}\in\mathbb{R}^{n\times d_{\text{model}}}\) is thereby obtained. Adding the embedding matrix of the token sequence and the positional encoding matrix yields the input matrix \(\boldsymbol{X}^{\text{in}}\) of embeddings with positional information:
\begin{equation}
\label{eq4}
\boldsymbol{X}^{\mathrm{in}} = \boldsymbol{X}^{\mathrm{emb}} + \boldsymbol{PE},\quad \boldsymbol{X}^{\mathrm{in}} \in \mathbb{R}^{n\times d_{\mathrm{model}}}
\end{equation}

Let \(\boldsymbol{x}_t^{\text{in}}\) be the \(t\)-th row vector of \(\boldsymbol{X}^{\text{in}}\), i.e., the \(d_{\text{model}}\)-dimensional vector characterizing the comprehensive information of a single token. The three row vectors \(\boldsymbol{q}_t\), \(\boldsymbol{k}_t\) and \(\boldsymbol{v}_t\) can be obtained via the following operations:
\begin{equation}
\label{eq5}
\begin{aligned}
\boldsymbol{q}_t &= \boldsymbol{x}_t^{\mathrm{in}} \boldsymbol{W}^{\boldsymbol{Q}},\quad \boldsymbol{W}^{\boldsymbol{Q}} \in \mathbb{R}^{d_{\mathrm{model}}\times d_k},\quad \boldsymbol{q}_t \in \mathbb{R}^{d_k}\\
\boldsymbol{k}_t &= \boldsymbol{x}_t^{\mathrm{in}} \boldsymbol{W}^{\boldsymbol{K}},\quad \boldsymbol{W}^{\boldsymbol{K}} \in \mathbb{R}^{d_{\mathrm{model}}\times d_k},\quad \boldsymbol{k}_t \in \mathbb{R}^{d_k}\\
\boldsymbol{v}_t &= \boldsymbol{x}_t^{\mathrm{in}} \boldsymbol{W}^{\boldsymbol{V}},\quad \boldsymbol{W}^{\boldsymbol{V}} \in \mathbb{R}^{d_{\mathrm{model}}\times d_v},\quad \boldsymbol{v}_t \in \mathbb{R}^{d_v}
\end{aligned}
\end{equation}

where \(\boldsymbol{W}^{\boldsymbol{Q}}\), \(\boldsymbol{W}^{\boldsymbol{K}}\) and \(\boldsymbol{W}^{\boldsymbol{V}}\) are trainable projection weight matrices. Stacking the computed results of all row vectors in \(\boldsymbol{X}^{\text{in}}\) row by row yields the corresponding matrices \(\boldsymbol{Q}\), \(\boldsymbol{K}\) and \(\boldsymbol{V}\):
\begin{equation}
\label{eq6}
\begin{aligned}
\boldsymbol{Q} &= \boldsymbol{X}^{\mathrm{in}} \boldsymbol{W}^{\boldsymbol{Q}},\quad \boldsymbol{Q} \in \mathbb{R}^{n\times d_k}\\
\boldsymbol{K} &= \boldsymbol{X}^{\mathrm{in}} \boldsymbol{W}^{\boldsymbol{K}},\quad \boldsymbol{K} \in \mathbb{R}^{n\times d_k}\\
\boldsymbol{V} &= \boldsymbol{X}^{\mathrm{in}} \boldsymbol{W}^{\boldsymbol{V}},\quad \boldsymbol{V} \in \mathbb{R}^{n\times d_v}
\end{aligned}
\end{equation}

The raw attention score matrix is thereby computed as follows:
\begin{equation}
\label{eq7}
\boldsymbol{S} = \boldsymbol{Q}\boldsymbol{K}^{\mathrm{T}}
\end{equation}

The element \(s_{l,m}\) in matrix \(\boldsymbol{S}\) denotes the raw dot product of the \(l\)-th row vector in \(\boldsymbol{Q}\) and the \(m\)-th row vector in \(\boldsymbol{K}\), characterizing their similarity:
\begin{equation}
\label{eq8}
s_{l,m} = \boldsymbol{q}_l \cdot \boldsymbol{k}_m^{\mathrm{T}}
\end{equation}

When \(d_k\) is large, the variance of dot-product results increases significantly, which may lead to vanishing gradients in subsequent computations. For this reason, the original paper further divides the raw attention score matrix \(\boldsymbol{S}\) by the scaling factor \(\sqrt{d_k}\) to obtain the scaled attention score matrix \(\boldsymbol{S}^{\text{scaled}}\), whose element is \(s^{\text{scaled}}_{l,m}\):
\begin{equation}
\label{eq9}
\begin{aligned}
\boldsymbol{S}^{\text{scaled}} &= \frac{\boldsymbol{S}}{\sqrt{d_k}} = \frac{\boldsymbol{Q}\boldsymbol{K}^{\mathrm{T}}}{\sqrt{d_k}} \\
s_{l,m}^{\text{scaled}} &= \frac{\boldsymbol{q}_l \cdot \boldsymbol{k}_m^{\mathrm{T}}}{\sqrt{d_k}}
\end{aligned}
\end{equation}

On this basis, softmax normalization is computed row by row for the scaled attention score matrix \(\boldsymbol{S}^{\text{scaled}}\), yielding the attention weight matrix \(\boldsymbol{S}^{\text{softmax}}\), whose element is \(s^{\text{softmax}}_{l,m}\):
\begin{equation}
\label{eq10}
\begin{gathered}
\boldsymbol{S}^{\text{softmax}} = \mathrm{softmax}\big(\boldsymbol{S}^{\text{scaled}}\big), \quad \boldsymbol{S}^{\text{softmax}} \in \mathbb{R}^{n\times n} \\[8pt]
s_{l,m}^{\text{softmax}} = \frac{\exp\big(s_{l,m}^{\text{scaled}}\big)}{\displaystyle\sum_{m'=0}^{n-1}\exp\big(s_{l,m'}^{\text{scaled}}\big)}
\end{gathered}
\end{equation}

Finally, the attention weight matrix \(\boldsymbol{S}^{\text{softmax}}\) is multiplied by the matrix \(\boldsymbol{V}\) to obtain the scaled dot-product attention output matrix \(\boldsymbol{Z}\) that encodes the contextual interaction information between token embeddings, with row vector \(\boldsymbol{z}_t\):
\begin{equation}
\label{eq11}
\begin{gathered}
\boldsymbol{Z} = \boldsymbol{S}^{\text{softmax}} \boldsymbol{V}, \quad \boldsymbol{S}^{\text{softmax}} \in \mathbb{R}^{n\times n},\quad \boldsymbol{Z} \in \mathbb{R}^{n\times d_v} \\[8pt]
\boldsymbol{z}_t = \boldsymbol{s}_t^{\text{softmax}} \boldsymbol{V}
\end{gathered}
\end{equation}

From the above computations, the formula for scaled dot-product attention proposed in the original paper can be derived:
\begin{equation}
\label{eq12}
\mathrm{Attention}(\boldsymbol{Q},\boldsymbol{K},\boldsymbol{V}) = \mathrm{softmax}\left( \frac{\boldsymbol{Q}\boldsymbol{K}^{\mathrm{T}}}{\sqrt{d_k}} \right)\boldsymbol{V}
\end{equation}

Based on the foregoing review, the computational logic and core steps of the classical self-attention mechanism can be summarized. First, the range of the token sequence for analysis and computation is defined. All tokens within this range are mapped to corresponding token embeddings, and positional encoding information is incorporated to yield the input matrix \(\boldsymbol{X}^{\text{in}}\). It should be noted that the token embeddings at this stage are completely independent of each other, so \(\boldsymbol{X}^{\text{in}}\) has not yet encoded the interaction information between tokens. Next, matrices \(\boldsymbol{Q}\), \(\boldsymbol{K}\) and \(\boldsymbol{V}\) are computed, and scaled dot-product operations are performed on \(\boldsymbol{Q}\) and \(\boldsymbol{K}\) to obtain the raw attention score matrix \(\boldsymbol{S}\), whose elements correspond to the raw scores for any pair of token embeddings. Subsequently, scaling and softmax normalization are applied to \(\boldsymbol{S}\) to yield the attention distribution of any token embedding over all token embeddings in the sequence (including itself), namely the attention weight matrix \(\boldsymbol{S}^{\text{softmax}}\). Finally, \(\boldsymbol{S}^{\text{softmax}}\) is multiplied by matrix \(\boldsymbol{V}\) to yield the scaled dot-product attention output \(\boldsymbol{Z}\), which incorporates positional information and encodes pairwise interactions among tokens.

\begin{figure}[htbp]
\centering
\includegraphics[width=0.85\textwidth]{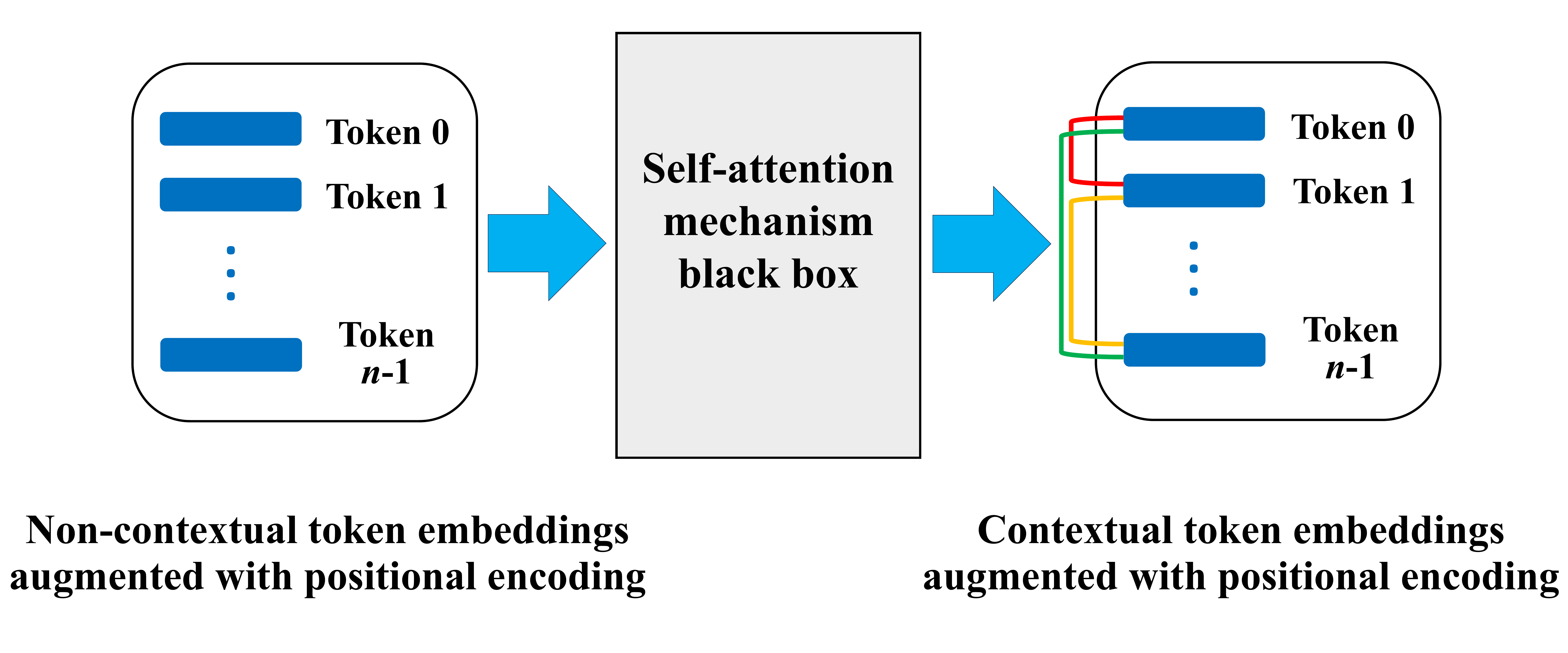}
\vspace{-0.5em}
\caption{Self-Attention Mechanism Black Box}
\vspace{-0.8em}
\label{fig1}
\end{figure}

The above computational logic indicates that the self-attention mechanism can essentially be regarded as a black box for encoding the degree of interactions between tokens in a sequence, as illustrated in Figure~\ref{fig1}. It leverages matrix operations to process raw non-contextual token embeddings augmented with positional encoding. By modeling pairwise interactions among all tokens, including self-interactions, it transforms these static embeddings into contextualized representations that capture global context. Specifically, each output embedding is obtained as a weighted sum over all input embeddings in the sequence via softmax-normalized attention weights. From a functionalist perspective, this implies that if another similar mechanism can realize self-attention encoding of token interactions, it can replace the classical self-attention mechanism and serve as the core engine for natural language processing.

However, the architecture of the classical self-attention mechanism has two inherent drawbacks that are difficult to mitigate. First, discrete matrix computations lead to the loss of continuous semantic features. In reality, natural language is a semantic information flow that evolves continuously over time. For engineering considerations, the classical self-attention mechanism truncates this semantic flow into discrete matrix forms and computes attention scores via explicit matrix operations, breaking the inherent continuous semantic structure of language. Although tokenization is an indispensable preprocessing step for text modeling, this computational scheme erodes semantic integrity and inevitably sacrifices the implicit continuity that enables smooth transitions during the sequential evolution of meaning, thereby disrupting the native semantic context of text. The model can only reconstruct global semantics through explicit weight matching between discrete features. Second, numerous complex matrix operations incur substantial computational overhead. The mechanism relies on multiple matrix multiplications to perform feature projection and global correlation modeling, consuming considerable computational and storage resources, with both time and space complexity of \(\mathcal{O}(n^2 d_{\text{model}})\). As the token sequence length \(n\) increases, the computational complexity grows sharply on the order of \(n^2 d_{\text{model}}\). Even if a sliding window strategy is introduced to constrain the interaction range and reduce computational scale, the model still needs to explicitly compute attention scores for all token combinations within each window, resulting in high computational overhead.

To address the above two drawbacks of the classical self-attention mechanism, this work replaces the mechanism with an FC-QBM architecture and constructs a quantum language model on this basis. It will be subsequently demonstrated that FC-QBM can fully encode token positions and interactions. Functionally equivalent to the classical self-attention mechanism, this architecture is expected to reduce resource consumption by leveraging the parallelism and entanglement properties of quantum computing, thus enabling it to act as a core engine for natural language processing tasks.

Most notably, unlike the classical self-attention mechanism that explicitly decomposes continuous semantic flows into discrete matrices to compute attention scores, FC-QBM can implicitly encode positional information and capture pairwise interactions entirely through the unitary evolution of semantic quantum states. This approach not only circumvents the disruption to the continuous semantic structure of natural language caused by the discretized modeling of classical self-attention, but also maximally preserves the implicit continuous information that exhibits smooth transitions during the sequential evolution of semantics. In addition, it fundamentally reduces the computational overhead introduced by extensive redundant computations. Moreover, the probability distribution yielded by quantum measurements naturally satisfies the normalization constraint, eliminating the need for the softmax operation and further reducing computational complexity.

\subsection{Computational Principle of Self-Attention Mechanism Based on Fully Connected Quantum Boltzmann Machine}

This work constructs a computational module for the self-attention mechanism based on FC-QBM. It should be emphasized that, although this section derives in detail the computational principle of the self-attention mechanism built upon FC-QBM, it does not imply that explicit computation must be performed step by step strictly following the derivation process. In fact, the advantage of quantum computing is that once the corresponding quantum circuit is designed according to established rules, the system can spontaneously and implicitly yield the result via unitary evolution of the initial state. Let the real vector of the token sequence processed by quantum amplitude encoding be \(\boldsymbol{a}=[a_0\ a_1\ \dots\ a_{2^N-1}]\), where \(N\) denotes the number of qubits in the quantum circuit. This vector is then encoded into the quantum circuit as the amplitudes of basis states to prepare the initial state \(|\psi_0\rangle\):
\begin {equation}
\label {eq13}
\left|\psi_0\right\rangle=\sum_{\boldsymbol {r}=0}^{2^N-1} a_{\boldsymbol {r}} \left|\boldsymbol {r}\right\rangle,\quad a_{\boldsymbol {r}} \ge 0\;\,\text { and }\sum_{\boldsymbol {r}=0}^{2^N-1} a_{\boldsymbol {r}}^2 = 1
\end {equation}

For simplicity, Equation~\ref{eq13} adopts the shorthand notation \(\boldsymbol{r}\) for the binary string \(r_0r_1\cdots r_{N-1}\) corresponding to the measurement outcomes of individual qubits, and \(|\boldsymbol{r}\rangle\) denotes a basis vector from the space spanned by \(N\) qubits. FC-QBM adapted to the quantum circuit is defined by Hamiltonian \(H\):
\begin{equation}
\label{eq14}
H=\sum_{i=0}^{N-1} b_i \sigma_{\mathrm{Z}}^i + \sum_{i=0}^{N-1}\sum_{j=i+1}^{N-1} w_{ij}\sigma_{\mathrm{Z}}^i\sigma_{\mathrm{Z}}^j
\end{equation}

where \(b_i\) denotes the bias of the \(i\)-th unit, and \(w_{ij}\) denotes the connection weight between the \(i\)-th unit and the \(j\)-th unit, both of which are trainable parameters. \(\sigma_\mathrm{Z}^i\) denotes the Pauli-Z gate acting solely on the \(i\)-th qubit, and \(\sigma_\mathrm{Z}^i\sigma_\mathrm{Z}^j\) denotes the tensor product of two Pauli-Z gates acting respectively on the \(i\)-th qubit and the \(j\)-th qubit:
\begin{equation}
\label{eq15}
\begin{gathered}
\sigma_{\mathrm{Z}}^i = \mathrm{I}^{\otimes i} \otimes \sigma_{\mathrm{Z}} \otimes \mathrm{I}^{\otimes (N-i-1)} \\
\sigma_{\mathrm{Z}}^i\sigma_{\mathrm{Z}}^j = \mathrm{I}^{\otimes i} \otimes \sigma_{\mathrm{Z}} \otimes \mathrm{I}^{\otimes (j-i-1)} \otimes \sigma_{\mathrm{Z}} \otimes \mathrm{I}^{\otimes (N-j-1)}
\end{gathered}
\end{equation}

where \(\mathrm{I}\) denotes the identity gate. The Pauli-Z gate satisfies the following properties:
\begin{equation}
\label{eq16}
\sigma_{\mathrm{Z}} \ket{0} = \ket{0},\quad \sigma_{\mathrm{Z}} \ket{1} = -\ket{1}
\end{equation}

The classical FC-BM energy function \(\mathit{EF}\) corresponding to \(H\) is:
\begin{equation}
\label{eq17}
EF = \sum_{i=0}^{N-1} b_i unit_i + \sum_{i=0}^{N-1}\sum_{j=i+1}^{N-1} w_{ij} unit_i unit_j
\end{equation}

In Equation~\ref{eq17}, \(\mathit{unit}_i\) and \(\mathit{unit}_j\) denote the values of the \(i\)-th unit and the \(j\)-th unit, respectively, with \(\mathit{unit}_i\in\{-1, 1\}\) and \(\mathit{unit}_j\in\{-1, 1\}\). The corresponding probability distribution function \(P\) and partition function \(Z\) are given by:
\begin{equation}
\label{eq18}
\begin{gathered}
P = \frac{1}{Z}\exp\left(-\left(\sum_{i=0}^{N-1} b_i unit_i + \sum_{i=0}^{N-1}\sum_{j=i+1}^{N-1} w_{ij} unit_i unit_j\right)\right) \\
Z = \sum_{unit_i,unit_j\in\{-1,1\}} \exp\left(-\left(\sum_{i=0}^{N-1} b_i unit_i + \sum_{i=0}^{N-1}\sum_{j=i+1}^{N-1} w_{ij} unit_i unit_j\right)\right)
\end{gathered}
\end{equation}

Based on the properties of the Pauli-Z gate given in Eqation~\ref{eq16}, the result obtained by applying the Hamiltonian \(H\) to an arbitrary basis vector is:
\begin{equation}
\label{eq19}
\begin{aligned}
H|\boldsymbol{r}\rangle &= \left(\sum_{i=0}^{N-1} b_i \sigma_{\mathrm{Z}}^i + \sum_{i=0}^{N-1}\sum_{j=i+1}^{N-1} w_{ij} \sigma_{\mathrm{Z}}^i \sigma_{\mathrm{Z}}^j\right)|\boldsymbol{r}\rangle \\
&= \left(\sum_{i=0}^{N-1} b_i (-1)^{r_i} + \sum_{i=0}^{N-1}\sum_{j=i+1}^{N-1} w_{ij} (-1)^{r_i+r_j}\right)|\boldsymbol{r}\rangle \\
&= \omega(\boldsymbol{r})|\boldsymbol{r}\rangle
\end{aligned}
\end{equation}

The unitary evolution operator corresponding to \(H\) is \(U_H\):
\begin{equation}
\label{eq20}
U_H = \mathrm{e}^{-\mathrm{i}Ht} = \exp\left(-\mathrm{i}\left(\sum_{i=0}^{N-1} b_i \sigma_{\mathrm{Z}}^i + \sum_{i=0}^{N-1}\sum_{j=i+1}^{N-1} w_{ij} \sigma_{\mathrm{Z}}^i \sigma_{\mathrm{Z}}^j\right)t\right)
\end{equation}

Let \(\varphi(\boldsymbol{r})=\omega(\boldsymbol{r})t\). According to the properties of the Hamiltonian eigenvalue equation, applying the unitary evolution operator obtained from Equation~\ref{eq20} to the initial state \(|\psi_0\rangle\) results in the intermediate state \(|\psi_1\rangle\):
\begin{equation}
\label{eq21}
|\psi_1\rangle = U_H|\psi_0\rangle = \sum_{\boldsymbol{r}=0}^{2^N-1} a_{\boldsymbol{r}} \mathrm{e}^{-\mathrm{i}Ht}|\boldsymbol{r}\rangle = \sum_{\boldsymbol{r}=0}^{2^N-1} a_{\boldsymbol{r}} \mathrm{e}^{-\mathrm{i}\omega(\boldsymbol{r})t}|\boldsymbol{r}\rangle = \sum_{\boldsymbol{r}=0}^{2^N-1} a_{\boldsymbol{r}} \mathrm{e}^{-\mathrm{i}\varphi(\boldsymbol{r})}|\boldsymbol{r}\rangle
\end{equation}

At the end of the circuit, a Y-axis rotation gate \(R_\mathrm{Y}(2\theta_i)\) is placed on each qubit, and the corresponding unitary evolution operator \(U_R\) is:
\begin{equation}
\label{eq22}
U_R = \bigotimes_{i=0}^{N-1} R_\mathrm{Y}(2\theta_i)
\end{equation}

Subsequently, projective measurements are performed on the measurement basis vector \(|\boldsymbol{l}\rangle\) composed of \(N\) qubits, where \(\boldsymbol{l}\) is the shorthand notation of the binary string \(l_0l_1\cdots l_{N-1}\). After applying \(U_R\) to the basis vector \(|\boldsymbol{r}\rangle\) used to expand the intermediate state \(|\psi_1\rangle\), the probability amplitude \(\beta_{\boldsymbol{l},\boldsymbol{r}}\) projected onto the measurement basis vector \(|\boldsymbol{l}\rangle\) is:
\begin{equation}
\label{eq23}
\beta_{\boldsymbol{l},\boldsymbol{r}} = \langle \boldsymbol{l}|U_R|\boldsymbol{r}\rangle = \langle \boldsymbol{l}|\bigotimes_{i=0}^{N-1} R_\mathrm{Y}(2\theta_i)|\boldsymbol{r}\rangle = \prod_{i=0}^{N-1} \langle l_i|R_\mathrm{Y}(2\theta_i)|r_i\rangle
\end{equation}

In Equation~\ref{eq23}, the matrix form of \(R_\mathrm{Y}(2\theta_i)\) and its corresponding matrix elements are:
\begin{equation}
\label{eq24}
\begin{gathered}
R_\mathrm{Y}(2\theta_i)=
\begin{bmatrix}
\cos\theta_i & -\sin\theta_i \\
\sin\theta_i & \cos\theta_i
\end{bmatrix} \\[6pt]
\langle 0|R_\mathrm{Y}(2\theta_i)|0\rangle=\cos\theta_i \quad \langle 0|R_\mathrm{Y}(2\theta_i)|1\rangle=-\sin\theta_i \\
\langle 1|R_\mathrm{Y}(2\theta_i)|0\rangle=\sin\theta_i \quad \langle 1|R_\mathrm{Y}(2\theta_i)|1\rangle=\cos\theta_i
\end{gathered}
\end{equation}

The final state \(|\psi_2\rangle\) after applying \(U_R\) to the intermediate state \(|\psi_1\rangle\) is:
\begin{equation}
\label{eq25}
|\psi_2\rangle = U_R |\psi_1\rangle = \sum_{\boldsymbol{r}=0}^{2^N-1} a_{\boldsymbol{r}} \mathrm{e}^{-\mathrm{i}\varphi(\boldsymbol{r})} U_R |\boldsymbol{r}\rangle
\end{equation}

Projecting the final state \(|\psi_2\rangle\) onto the measurement basis vector \(|\boldsymbol{l}\rangle\) yields the corresponding amplitude \(c_{\boldsymbol{l}}\):
\begin{equation}
\label{eq26}
c_{\boldsymbol{l}} = \langle \boldsymbol{l} | \psi_2 \rangle = \langle \boldsymbol{l} | U_R | \psi_1 \rangle
= \sum_{\boldsymbol{r}=0}^{2^N-1} a_{\boldsymbol{r}} \mathrm{e}^{-\mathrm{i}\varphi(\boldsymbol{r})} \langle \boldsymbol{l} | U_R | \boldsymbol{r} \rangle
= \sum_{\boldsymbol{r}=0}^{2^N-1} \beta_{\boldsymbol{l},\boldsymbol{r}} a_{\boldsymbol{r}} \mathrm{e}^{-\mathrm{i}\varphi(\boldsymbol{r})}
\end{equation}

At this point, the probability of measuring an arbitrary basis vector \(|\boldsymbol{l}\rangle\) is:
\begin{equation}
\label{eq27}
\begin{aligned}
p_{\boldsymbol{l}} &= \left| c_{\boldsymbol{l}} \right|^2 \\
&= c_{\boldsymbol{l}} \cdot c_{\boldsymbol{l}}^* \\
&= \left( \sum_{\boldsymbol{r}=0}^{2^N-1} \beta_{\boldsymbol{l},\boldsymbol{r}} a_{\boldsymbol{r}} \mathrm{e}^{-\mathrm{i}\varphi(\boldsymbol{r})} \right)
\left( \sum_{\boldsymbol{r}'=0}^{2^N-1} \beta_{\boldsymbol{l},\boldsymbol{r}'} a_{\boldsymbol{r}'} \mathrm{e}^{\mathrm{i}\varphi(\boldsymbol{r}')} \right) \\
&= \sum_{\boldsymbol{r}=0}^{2^N-1} \sum_{\boldsymbol{r}'=0}^{2^N-1} \beta_{\boldsymbol{l},\boldsymbol{r}} \beta_{\boldsymbol{l},\boldsymbol{r}'} a_{\boldsymbol{r}} a_{\boldsymbol{r}'} \mathrm{e}^{\mathrm{i}\left(\varphi(\boldsymbol{r}')-\varphi(\boldsymbol{r})\right)}
\end{aligned}
\end{equation}

In the expression for the probability \(p_{\boldsymbol{l}}\) from Equation~\ref{eq27}, when \(\boldsymbol{r}=\boldsymbol{r}'\), the following result holds:\hfill\mbox{}
\[
\begin{gathered}
\varphi(\boldsymbol{r}) = \varphi(\boldsymbol{r}') \implies \mathrm{e}^{\mathrm{i}\left(\varphi(\boldsymbol{r}')-\varphi(\boldsymbol{r})\right)} = \mathrm{e}^{0} = 1 \\
\beta_{\boldsymbol{l},\boldsymbol{r}} = \beta_{\boldsymbol{l},\boldsymbol{r}'} \\
a_{\boldsymbol{r}} = a_{\boldsymbol{r}'}
\end{gathered}
\]

Therefore, the corresponding term in the expression is:
\[
\sum_{\boldsymbol{r}=0}^{2^N-1} \beta_{\boldsymbol{l},\boldsymbol{r}}^2 a_{\boldsymbol{r}}^2
\]

When \(\boldsymbol{r}\neq\boldsymbol{r}'\), the corresponding term in the expression is:
\[
\begin{aligned}
&\sum_{\boldsymbol{r}=0}^{2^N-1}\sum_{\boldsymbol{r}'=0}^{2^N-1}
\Big(\beta_{\boldsymbol{l},\boldsymbol{r}}\beta_{\boldsymbol{l},\boldsymbol{r}'}a_{\boldsymbol{r}}a_{\boldsymbol{r}'}\mathrm{e}^{\mathrm{i}\big(\varphi(\boldsymbol{r}')-\varphi(\boldsymbol{r})\big)}
+\beta_{\boldsymbol{l},\boldsymbol{r}'}\beta_{\boldsymbol{l},\boldsymbol{r}}a_{\boldsymbol{r}'}a_{\boldsymbol{r}}\mathrm{e}^{\mathrm{i}\big(\varphi(\boldsymbol{r})-\varphi(\boldsymbol{r}')\big)}\Big) \\
=&\sum_{\boldsymbol{r}=0}^{2^N-1}\sum_{\boldsymbol{r}'=0}^{2^N-1}
\beta_{\boldsymbol{l},\boldsymbol{r}}\beta_{\boldsymbol{l},\boldsymbol{r}'}a_{\boldsymbol{r}}a_{\boldsymbol{r}'}
\Big(\mathrm{e}^{\mathrm{i}\big(\varphi(\boldsymbol{r}')-\varphi(\boldsymbol{r})\big)}
+\mathrm{e}^{\mathrm{i}\big(\varphi(\boldsymbol{r})-\varphi(\boldsymbol{r}')\big)}\Big) \\
=&\sum_{\boldsymbol{r}=0}^{2^N-1}\sum_{\boldsymbol{r}'=0}^{2^N-1}
\beta_{\boldsymbol{l},\boldsymbol{r}}\beta_{\boldsymbol{l},\boldsymbol{r}'}a_{\boldsymbol{r}}a_{\boldsymbol{r}'}
\Big(\mathrm{e}^{\mathrm{i}\big(\varphi(\boldsymbol{r}')-\varphi(\boldsymbol{r})\big)}
+\mathrm{e}^{-\mathrm{i}\big(\varphi(\boldsymbol{r}')-\varphi(\boldsymbol{r})\big)}\Big)
\end{aligned}
\]

The expression inside the parentheses contains two exponential terms. According to Euler's formula:
\[
\begin{aligned}
&\phantom{=}\:\:\, \mathrm{e}^{\mathrm{i}\big(\varphi(\boldsymbol{r}')-\varphi(\boldsymbol{r})\big)}+\mathrm{e}^{-\mathrm{i}\big(\varphi(\boldsymbol{r}')-\varphi(\boldsymbol{r})\big)} \\
&= \cos\big(\varphi(\boldsymbol{r}')-\varphi(\boldsymbol{r})\big)+\mathrm{i}\cdot\sin\big(\varphi(\boldsymbol{r}')-\varphi(\boldsymbol{r})\big)+\cos\big(-\big(\varphi(\boldsymbol{r}')-\varphi(\boldsymbol{r})\big)\big)+\mathrm{i}\cdot\sin\big(-\big(\varphi(\boldsymbol{r}')-\varphi(\boldsymbol{r})\big)\big) \\
&= \cos\big(\varphi(\boldsymbol{r}')-\varphi(\boldsymbol{r})\big)+\mathrm{i}\cdot\sin\big(\varphi(\boldsymbol{r}')-\varphi(\boldsymbol{r})\big)+\cos\big(\varphi(\boldsymbol{r}')-\varphi(\boldsymbol{r})\big)-\mathrm{i}\cdot\sin\big(\varphi(\boldsymbol{r}')-\varphi(\boldsymbol{r})\big) \\
&= 2\cos\big(\varphi(\boldsymbol{r}')-\varphi(\boldsymbol{r})\big)
\end{aligned}
\]

Without loss of generality, to avoid redundant computations, let \(\boldsymbol{r}<\boldsymbol{r}'\). Thus, when \(\boldsymbol{r}\neq\boldsymbol{r}'\), the corresponding term in Equation~\ref{eq27} simplifies to:
\[
2\sum_{r=0}^{2^N-1}\sum_{r'=r+1}^{2^N-1}\beta_{l,\boldsymbol{r}}\beta_{l,\boldsymbol{r}'}a_{\boldsymbol{r}}a_{\boldsymbol{r}'}\cos\big(\varphi(\boldsymbol{r}')-\varphi(\boldsymbol{r})\big)
\]

The probability \(p_{\boldsymbol{l}}\) of measuring an arbitrary basis vector \(|\boldsymbol{l}\rangle\) is hence given by:
\begin{equation}
\label{eq28}
p_{\boldsymbol{l}}=\sum_{\boldsymbol{r}=0}^{2^N-1}\beta_{\boldsymbol{l},\boldsymbol{r}}^2 a_{\boldsymbol{r}}^2
+2\sum_{\boldsymbol{r}=0}^{2^N-1}\sum_{\boldsymbol{r}'=\boldsymbol{r}+1}^{2^N-1}\beta_{\boldsymbol{l},\boldsymbol{r}}\beta_{\boldsymbol{l},\boldsymbol{r}'}a_{\boldsymbol{r}}a_{\boldsymbol{r}'}\cos\big(\varphi(\boldsymbol{r}')-\varphi(\boldsymbol{r})\big)
\end{equation}

The corresponding probability vector is \(\boldsymbol{p} = [p_{0}\ p_{1}\ \dots\ p_{2^{N}-1}]\).

\subsection{Equivalence Between Fully Connected Quantum Boltzmann Machine Self-Attention Mechanism and Classical Self-Attention Mechanism}

From the above derivations, the quantum amplitude encoding scheme for tokens proposed in this work concatenates the token sequence into a non-negative \(2^{N}\)-dimensional real vector subject to the constraints on quantum amplitudes, where each vector element corresponds to the amplitude of a basis vector. Each token embedding has a dimension of \(d_{\text{model}}\), and hence every such token is represented using \(d_{\text{model}}\) basis vector amplitudes. After modulation by FC-QBM and application of the \(R_\mathrm{Y}(2\theta_i)\) rotation gates, the positional and interaction information of tokens is encoded into the measurement probabilities given by Equation~\ref{eq28}. These measurement probabilities are equivalent to the results obtained from the classical self-attention mechanism.

First, expand the intermediate state \(|\psi_1\rangle\) modulated by FC-QBM given in Equation~\ref{eq21}, yielding:
\begin{equation}
\label{eq29}
\begin{aligned}
|\psi_1\rangle &= a_{\boldsymbol {r}} \mathrm{e}^{-\mathrm{i}\omega(\boldsymbol{r})t} |\boldsymbol{r}\rangle \\
&= a_{\boldsymbol {r}} \cdot \exp\left(-\mathrm{i}\left( \sum_{i=0}^{N-1} b_i (-1)^{r_i} + \sum_{i=0}^{N-1}\sum_{j=i+1}^{N-1} w_{ij} (-1)^{r_i+r_j} \right) t \right) |\boldsymbol{r}\rangle \\
&= a_{\boldsymbol {r}} \cdot \exp\left(-\mathrm{i}\cdot t \sum_{i=0}^{N-1} b_i (-1)^{r_i} \right) \exp\left(-\mathrm{i}\cdot t \sum_{i=0}^{N-1}\sum_{j=i+1}^{N-1} w_{ij} (-1)^{r_i+r_j} \right) |\boldsymbol{r}\rangle
\end{aligned}
\end{equation}

Equation~\ref{eq29} decomposes the global phase factor of the intermediate state into two components: the first is the bias phase factor, and the second is the weight phase factor. On the one hand, single-qubit phase combinations from the first component encode token order, assign positional markers to all tokens in the sequence, and thereby realize positional encoding. On the other hand, phase combinations generated by arbitrary two-qubit interactions in the second component encode the pairwise interactions between tokens at the qubit level, laying a foundation for subsequent matching with the classical attention score computation. It can thus be seen that the intermediate state \(|\psi_1\rangle\) modulated by FC-QBM implicitly encodes the positional and interaction information of all token embeddings in the sequence via combinations of qubit phases. After further application of the \(R_\mathrm{Y}(2\theta_i)\) rotation gates, \(|\psi_1\rangle\) evolves into the final state \(|\psi_2\rangle\), yielding measurement probabilities encoding the positional information and interaction strength of tokens.

The core computation of the classical self-attention mechanism is the dot product between token embeddings. Geometrically, the dot product equals the product of the norms of the two vectors multiplied by the cosine of the angle between them. Equation~\ref{eq28} naturally embodies this intuitive geometric interpretation, and the dot product of any two basis vectors is given by:\hfill\mbox{}
\begin{equation}
\label{eq30}
\begin{cases}
a_{\boldsymbol{r}}^2 & \text{if } \boldsymbol{r}=\boldsymbol{r}' \\
a_{\boldsymbol{r}} a_{\boldsymbol{r}'} \cos\big(\varphi(\boldsymbol{r}')-\varphi(\boldsymbol{r})\big) & \text{if } \boldsymbol{r}<\boldsymbol{r}'
\end{cases}
\end{equation}

At this point, a single token embedding is represented by the amplitudes of \(d_{\text{model}}\) basis vectors. Therefore, the dot product between token embeddings can be decomposed into the definition described in Equation~\ref{eq30}, which simultaneously includes the initial real-amplitude terms \(a_{\boldsymbol{r}}\) and \(a_{\boldsymbol{r}}'\) as well as the cosine term \(\cos(\varphi(\boldsymbol{r}')-\varphi(\boldsymbol{r}))\). The real-amplitude terms correspond to the quantum-state amplitudes of the original token sequence used as the initial input. These amplitudes carry no global interaction information. In contrast, the cosine term represents the phase information modulated by FC-QBM, encoding the positional information and interaction strength of token embeddings. Mathematically, Equation~\ref{eq30} fully conforms to the geometric interpretation of the dot-product operation. On this basis, by leveraging parameter combinations for the \(R_Y(2\theta_i)\) rotation gates assigned to respective qubits, different values for the coefficients \(\beta_{\boldsymbol{l},\boldsymbol{r}}\) and \(\beta_{\boldsymbol{l},\boldsymbol{r}'}\) in Equation~\ref{eq28} are obtained, thereby enabling biased encoding of the dot-product components under each distinct basis vector. This mechanism avoids uniform representation and ensures that the dot-product components under each basis vector can be adaptively adjusted in accordance with token features.

The above steps can be briefly summarized as follows. First, FC-QBM encodes token positions and global pairwise interactions, with this information represented as phases. Then, the \(R_Y(2\theta_i)\) rotation gates extract differentiated information. In this manner, the encoding and computational functions of classical self-attention are fully realized. Most notably, the unitary evolution of semantic quantum states enables implicit attention computation by exploiting quantum parallelism and entanglement, removing the need to decompose continuous semantic streams into discrete matrices for explicit attention-weight computation. Furthermore, circuit measurement returns a probability distribution over basis vectors, which naturally satisfies normalization and requires no separate softmax step. The resulting probability vector encodes token positions and interactions, and can be directly applied to downstream generation and recognition tasks. In summary, by leveraging the strengths of quantum computing, the method can theoretically greatly reduce computational complexity, speed up model inference and overcome the inherent drawbacks of classical self-attention.

\section{Construction and Implementation of Fully Connected Quantum Boltzmann Machine Language Model}

\subsection{Initial State Preparation and Quantum Circuit Construction}

First, quantum amplitude encoding is performed on the token sequence.

To begin with, an embedding matrix \(\boldsymbol{E}\in\mathbb{R}^{V\times d_{\text{model}}}\) is trained from the text corpus, where \(V\) denotes the vocabulary size and \(d_{\text{model}}\) represents the dimension of word embeddings. Each row of \(\boldsymbol{E}\) corresponds to a \(d_{\text{model}}\)-dimensional word embedding \(\boldsymbol{e}(x_t)\). For the convenience of subsequent derivation, this vector is abbreviated as \(\boldsymbol{e}_t\). Given a token sequence \(x_0, x_1, \dots, x_{n-1}\) with a length of \(n\), the \(n\) word embeddings are strictly concatenated in token order to form an \((n \times d_{\text{model}})\)-dimensional vector \(\boldsymbol{TK}_0\):
\begin{equation}
\label{eq31}
\boldsymbol{TK}_0 = \big[\boldsymbol{e}_0\;\boldsymbol{e}_1\;\cdots\;\boldsymbol{e}_{n-1}\big]
= \big[\;tk_{0,0}\quad tk_{0,1}\quad \cdots\quad tk_{0,n\times d_{\text{model}}-1}\;\big]
\end{equation}

It can be seen from Equation~\ref{eq31} that after concatenation, every \(d_{\text{model}}\) elements in vector \(\boldsymbol{TK}_0\) represent one token from left to right. There are a total of \(n\times d_{\text{model}}\) elements, which can characterize \(n\) tokens. Subsequently, the minimum value among all elements of \(\boldsymbol{TK}_0\) is subtracted from these elements to make all vector elements non-negative, and a new vector \(\boldsymbol{TK}_1\) is obtained accordingly:
\begin{equation}
\label{eq32}
\begin{gathered}
\min\left(\boldsymbol{TK}_0\right) = \min_{0\le i \le n\times d_{\text{model}}-1} \big(tk_{0,i}\big),\quad tk_{1,i} = tk_{0,i} - \min\left(\boldsymbol{TK}_0\right) \\
\boldsymbol{TK}_1 = \big[\;tk_{1,0}\quad tk_{1,1}\quad \dots\quad tk_{1,n\times d_{\text{model}}-1}\;\big]
\end{gathered}
\end{equation}

Then, \(\boldsymbol{TK}_1\) is processed via L1 normalization to yield vector \(\boldsymbol{TK}_2\):
\begin{equation}
\label{eq33}
\begin{gathered}
tk_{2,i} = \frac{tk_{1,i}}{\displaystyle\sum_{i'=0}^{n\times d_{\text{model}}-1} tk_{1,i'}} \\
\boldsymbol{TK}_2 = \big[\;tk_{2,0}\quad tk_{2,1}\quad \dots\quad tk_{2,n\times d_{\text{model}}-1}\;\big]
\end{gathered}
\end{equation}

To satisfy the requirements for quantum state amplitudes, all elements in vector \(\boldsymbol{TK}_2\) are further square-rooted to yield vector \(\boldsymbol{TK}_{\text{sqr}}\):
\begin{equation}
\label{eq34}
\begin{gathered}
tk_{\text{sqr},i} = \sqrt{tk_{2,i}} \\
\boldsymbol{TK}_{\text{sqr}} = \big[\;tk_{\text{sqr},0}\quad tk_{\text{sqr},1}\quad \dots\quad tk_{\text{sqr},n\times d_{\text{model}}-1}\;\big]
\end{gathered}
\end{equation}

Let the number of qubits in the quantum circuit be \(N\), which corresponds to a \(2^N\)-dimensional Hilbert space. Therefore, the total capacity of amplitude encoding is \(2^N\). To encode all elements of \(\boldsymbol{TK}_{\text{sqr}}\), \(2^N\) must satisfy the following capacity constraint:
\begin{equation}
\label{eq35}
2^{N} \ge n\times d_{\text{model}}
\end{equation}

The minimum number of qubits \(N\) satisfying the above constraints is computed using the ceiling function:
\begin{equation}
\label{eq36}
N = \left\lceil \log_2 \left(n\times d_{\text{model}}\right) \right\rceil
\end{equation}

If the number of basis vectors slightly exceeds the number of required elements, the amplitudes corresponding to the extra basis vectors are padded with zeros, thereby yielding the \(2^N\)-dimensional vector \(\boldsymbol{TK}_{\text{amp}}\):
\begin{equation}
\label{eq37}
\begin{gathered}
tk_{\mathrm{amp},i}=
\begin{cases}
tk_{\mathrm{sqr},i} & \text{if}\quad 0\le i\le n\times d_{\text{model}}-1\\
0 & \text{if}\quad n\times d_{\text{model}}-1<i\le 2^{N}-1
\end{cases}\\[1em]
\boldsymbol{\mathit{TK}}_{\mathrm{amp}} = \bigl[\, tk_{\mathrm{amp},0}\ \ tk_{\mathrm{amp},1}\ \ \dots\ \ tk_{\mathrm{amp},2^{N}-1} \,\bigr]
\end{gathered}
\end{equation}

Finally, the vector \(\boldsymbol{TK}_{\text{amp}}\) computed via Equation~\ref{eq37} is assigned to vector \(\boldsymbol{a}\), thus completing the quantum amplitude encoding process for the token sequence:
\begin{equation}
\label{eq38}
\begin{gathered}
\mathit{a}_{\boldsymbol{\mathit{r}}} = tk_{\mathrm{amp},i},\quad \boldsymbol{\mathit{r}} = i\\
\boldsymbol{\mathit{a}} = \bigl[\, a_0\ \ a_1\ \ \dots\ \ a_{2^{N}-1} \,\bigr]
\end{gathered}
\end{equation}

where \(\boldsymbol{r}\) denotes the shorthand notation obtained by converting integer \(i\) into the binary string \(r_0r_1\cdots r_{N-1}\).

Second, an \(N\)-qubit quantum circuit is constructed, and the vector \(\boldsymbol{a}\) is encoded into the circuit as the amplitudes of basis vectors, thereby preparing the initial state \(|\psi_0\rangle\).

Third, FC-QBM and Y-axis rotation gates are constructed in the quantum circuit according to the detailed computation and derivation procedures presented above. Specifically, FC-QBM is implemented using \(R_\mathrm{Z}(2b_i)\), \(R_\mathrm{Z}(2w_{ij})\) and CNOT gates, and will be illustrated in detail with an 8-qubit model subsequently. The Y-axis rotation gates are realized via \(R_\mathrm{Y}(2\theta_i)\) gates. Here, \(b_i\), \(w_{ij}\) and \(\theta_i\) are trainable parameters. The number of \(w_{ij}\) is \(N(N-1)/2\). The number of \(b_i\) is \(N\), and the number of \(\theta_i\) is also \(N\).

\subsection{Construction of Classical Neural Network Layers}

To further predict the next token using the probability vector processed by the quantum circuit, this work constructs classical neural network layers following FC-QBM. Execution of the quantum circuit yields the final state \(|\psi_2\rangle\), while measurement produces the probability vector \(\boldsymbol{P} = [p_0 \ p_1 \ \dots \ p_{2N-1}]\). This vector, in turn, is fed into the classical neural network layers to predict the next token embedding. Specifically, the classical neural network layers consist of two linear layers with a Gaussian error linear unit (GELU) activation function\cite{hendrycks2016gaussian} inserted in between. The first classical linear layer fuses the features of \(\boldsymbol{P}\) and reduces its dimensionality, compressing it into an \(M\)-dimensional representation:
\begin{equation}
\label{eq39}
\begin{gathered}
c_{1,k} = \sum_{\boldsymbol{\mathit{l}}=0}^{2^{N}-1} w_{1,k,\boldsymbol{\mathit{l}}} p_{\boldsymbol{\mathit{l}}} + b_{1,k}\\
\boldsymbol{\mathit{C}}_1 = \bigl[\, c_{1,0}\ \ c_{1,1}\ \ \dots\ \ c_{1,M-1} \,\bigr]
\end{gathered}
\end{equation}

The bias \(b_{1,k}\) and the weight \(w_{1,k,\boldsymbol{l}}\) in Equation~\ref{eq39} are trainable parameters, where \(\boldsymbol{C}_1\) denotes the output vector processed by the first classical linear layer. The number of \(b_{1,k}\) is \(M\), and the number of \(w_{1,k,\boldsymbol{l}}\) is \(M\cdot 2^N\). Subsequently, elements of vector \(\boldsymbol{C}_1\) undergo nonlinear transformation via the GELU activation function, yielding vector \(\boldsymbol{C}_\text{GELU}\):
\begin{equation}
\label{eq40}
\begin{gathered}
c_{\mathrm{GELU},k} = c_{1,k} \cdot \Phi\left(c_{1,k}\right)\\
\boldsymbol{\mathit{C}}_{\mathrm{GELU}} = \bigl[\, c_{\mathrm{GELU},0}\ \ c_{\mathrm{GELU},1}\ \ \dots\ \ c_{\mathrm{GELU},M-1} \,\bigr]
\end{gathered}
\end{equation}

where \(\Phi(c_{1,k})\) denotes the cumulative distribution function for the standard normal distribution:\hfill\mbox{}
\begin{equation}
\label{eq41}
\Phi\left(c_{1,k}\right) = \frac{1}{\sqrt{2\pi}} \int_{-\infty}^{c_{1,k}} \exp\left(-\frac{x^2}{2}\right) dx
\end{equation}

Finally, vector \(\boldsymbol{C}_\text{GELU}\) is fed into the second classical linear layer to output the \(d_\text{model}\)-dimensional predicted token embedding \(\hat{\boldsymbol{e}}_n\):
\begin{equation}
\label{eq42}
\begin{gathered}
c_{2,s} = \sum_{k=0}^{M-1} w_{2,s,k} c_{\mathrm{GELU},k} + b_{2,s}\\[4pt]
\hat{\boldsymbol{e}}_n = \bigl[\, \hat{e}_{n,0}\ \ \hat{e}_{n,1}\ \ \dots\ \ \hat{e}_{n,d_{\mathrm{model}}-1} \,\bigr]
= \bigl[\, c_{2,0}\ \ c_{2,1}\ \ \dots\ \ c_{2,d_{\mathrm{model}}-1} \,\bigr]
\end{gathered}
\end{equation}

The bias \(b_{2,s}\) and the weight \(w_{2,s,k}\) in Equation~\ref{eq42} are trainable parameters. The number of \(b_{2,s}\) is \(d_\text{model}\), and the number of \(w_{2,s,k}\) is \(d_\text{model}\cdot M\).

\subsection{Model Training}

Combined with the foregoing analysis, the language model constructed in this work contains three groups of parameters. The first group consists of the FC-QBM circuit parameters \(b_i\), \(w_{ij}\) and \(\theta_i\), where the number of \(w_{ij}\) is \(N(N-1)/2\), and the numbers of \(b_i\) and \(\theta_i\) are each \(N\). The second group consists of the parameters \(b_{1,k}\) and \(w_{1,k,\boldsymbol{l}}\) of the first classical linear layer, where the number of \(b_{1,k}\) is \(M\), and the number of \(w_{1,k,\boldsymbol{l}}\) is \(M\cdot 2^N\). The third group consists of the parameters \(b_{2,s}\) and \(w_{2,s,k}\) of the second classical linear layer, where the number of \(b_{2,s}\) is \(d_\text{model}\), and the number of \(w_{2,s,k}\) is \(d_\text{model}\cdot M\). In practice, the corpus and the pre-trained embedding matrix are used to update model parameters through rounds of iterative optimization.

The training process of FC-BM is generally divided into two stages: positive-phase energy minimization and negative-phase contrastive divergence learning. Compared with classical FC-BM, FC-QBM constructed in this work, which acts as the core engine of the quantum language model, possesses unique features. First, FC-QBM is built by means of quantum circuits and various rotation gate operations. Second, classical neural network layers need to be executed after running FC-QBM. Accordingly, while following the standard training procedures of FC-BM, this work uses the backpropagation method to update the model parameters\cite{rumelhart1986learning}. Specifically, the forward propagation of the model is first used to simulate the positive-phase energy minimization of FC-BM. During forward propagation, the three groups of parameters are initially assigned random values, and then the model runs layer by layer from the quantum layer to the classical layers to obtain the predicted next token embedding. Subsequently, parameter gradients are obtained via model backpropagation for updating, so as to simulate negative-phase contrastive divergence learning.

Regarding the selection of the loss function, this work adopts the mean squared error (MSE) to measure the difference between predicted and ground-truth token embeddings. Mathematically, let the predicted token embedding generated by the model be \(\hat{\boldsymbol{e}}_n\), and the ground-truth token embedding be \(\boldsymbol{e}_n\). The corresponding loss function \(L\) is defined as:
\begin{equation}
\label{eq43}
\begin{gathered}
\hat{\boldsymbol{e}}_n = \bigl[\, \hat{e}_{n,0}\ \ \hat{e}_{n,1}\ \ \dots\ \ \hat{e}_{n,d_{\mathrm{model}}-1} \,\bigr],\quad
\boldsymbol{e}_n = \bigl[\, e_{n,0}\ \ e_{n,1}\ \ \dots\ \ e_{n,d_{\mathrm{model}}-1} \,\bigr]\\[6pt]
L = \frac{1}{d_{\mathrm{model}}}\sum_{s=0}^{d_{\mathrm{model}}-1} \left(\hat{e}_{n,s} - e_{n,s}\right)^2
\end{gathered}
\end{equation}

\subsubsection{Parameter Updates for Second Classical Linear Layer}

According to the computational rules of backpropagation, the gradients of the loss function with respect to the parameters \(b_{2,s}\) and \(w_{2,s,k}\) of the second classical linear layer at the output end are first computed, followed by the update of the parameters. The corresponding computational formulas are as follows:
\begin{equation}
\label{eq44}
\begin{gathered}
\frac{\partial L}{\partial b_{2,s}} = \frac{\partial L}{\partial \hat{e}_{n,s}} \cdot \frac{\partial \hat{e}_{n,s}}{\partial b_{2,s}},\quad
\frac{\partial L}{\partial w_{2,s,k}} = \frac{\partial L}{\partial \hat{e}_{n,s}} \cdot \frac{\partial \hat{e}_{n,s}}{\partial w_{2,s,k}}\\[4pt]
s = 0,\,1,\,\dots,\,d_{\mathrm{model}}-1,\\
k = 0,\,1,\,\dots,\,M-1
\end{gathered}
\end{equation}

According to Equation~\ref{eq42} and Equation~\ref{eq43}, the results for all terms in Equation~\ref{eq44} are:\hfill\mbox{}
\begin{equation}
\label{eq45}
\begin{gathered}
\frac{\partial L}{\partial \hat{e}_{n,s}}
= \frac{\partial \displaystyle\frac{1}{d_{\mathrm{model}}}\sum_{s=0}^{d_{\mathrm{model}}-1}\left(\hat{e}_{n,s}-e_{n,s}\right)^2}{\partial \hat{e}_{n,s}}
= \frac{2}{d_{\mathrm{model}}}\left(\hat{e}_{n,s}-e_{n,s}\right)=\delta_{n,s}\\[8pt]
\frac{\partial \hat{e}_{n,s}}{\partial b_{2,s}}=\frac{\partial c_{2,s}}{\partial b_{2,s}}=1\\[8pt]
\frac{\partial \hat{e}_{n,s}}{\partial w_{2,s,k}}=\frac{\partial c_{2,s}}{\partial w_{2,s,k}}=c_{\mathrm{GELU},k}
\end{gathered}
\end{equation}

Therefore, the final gradients of the loss function with respect to \(b_{2,s}\) and \(w_{2,s,k}\) are:
\begin{equation}
\label{eq46}
\frac{\partial L}{\partial b_{2,s}} = \delta_{n,s},\quad
\frac{\partial L}{\partial w_{2,s,k}} = \delta_{n,s} \cdot c_{\mathrm{GELU},k}
\end{equation}

The corresponding parameter update formulas are:
\begin{equation}
\label{eq47}
b_{2,s}^{(t+1)} \leftarrow b_{2,s}^{(t)} - \eta_2 \frac{\partial L}{\partial b_{2,s}},\quad
w_{2,s,k}^{(t+1)} \leftarrow w_{2,s,k}^{(t)} - \eta_2 \frac{\partial L}{\partial w_{2,s,k}}
\end{equation}

where \(\eta_2\) denotes the learning rate of the second classical linear layer.

\subsubsection{Parameter Updates for First Classical Linear Layer}

Next, the gradients of the loss function with respect to the parameters \(b_{1,k}\) and \(w_{1,k,\boldsymbol{l}}\) of the first classical linear layer are computed, followed by the update of the parameters. The corresponding computational formulas are as follows:
\begin{equation}
\label{eq48}
\begin{gathered}
\frac{\partial L}{\partial b_{1,k}}
= \left( \sum_{s=0}^{d_{\mathrm{model}}-1} \frac{\partial L}{\partial \hat{e}_{n,s}} \cdot \frac{\partial \hat{e}_{n,s}}{\partial c_{\mathrm{GELU},k}} \right)
\cdot \frac{\partial c_{\mathrm{GELU},k}}{\partial c_{1,k}}
\cdot \frac{\partial c_{1,k}}{\partial b_{1,k}}\\[10pt]
\frac{\partial L}{\partial w_{1,k,\boldsymbol{l}}}
= \left( \sum_{s=0}^{d_{\mathrm{model}}-1} \frac{\partial L}{\partial \hat{e}_{n,s}} \cdot \frac{\partial \hat{e}_{n,s}}{\partial c_{\mathrm{GELU},k}} \right)
\cdot \frac{\partial c_{\mathrm{GELU},k}}{\partial c_{1,k}}
\cdot \frac{\partial c_{1,k}}{\partial w_{1,k,\boldsymbol{l}}}\\[6pt]
k = 0,\,1,\,\dots,\,M-1\\
\boldsymbol{l} = 0,\,1,\,\dots,\,2^N -1
\end{gathered}
\end{equation}

The results for all terms in Equation~\ref{eq48} are:
\begin{equation}
\label{eq49}
\begin{gathered}
\sum_{s=0}^{d_{\mathrm{model}}-1} \frac{\partial L}{\partial \hat{e}_{n,s}} \cdot \frac{\partial \hat{e}_{n,s}}{\partial c_{\mathrm{GELU},k}}
= \sum_{s=0}^{d_{\mathrm{model}}-1} \delta_{n,s} \cdot \frac{\partial c_{2,s}}{\partial c_{\mathrm{GELU},k}}
= \sum_{s=0}^{d_{\mathrm{model}}-1} \delta_{n,s} \cdot w_{2,s,k}\\[10pt]
\frac{\partial c_{\mathrm{GELU},k}}{\partial c_{1,k}}
= \Phi\left(c_{1,k}\right)+c_{1,k}\Phi'\left(c_{1,k}\right)
= \frac{1}{\sqrt{2\pi}} \left( \int_{-\infty}^{c_{1,k}} \exp\left(-\frac{x^2}{2}\right) dx + c_{1,k}\exp\left(-\frac{c_{1,k}^2}{2}\right) \right)\\[10pt]
\frac{\partial c_{1,k}}{\partial b_{1,k}} = 1\\[10pt]
\frac{\partial c_{1,k}}{\partial w_{1,k,\boldsymbol{l}}} = p_{\boldsymbol{l}}
\end{gathered}
\end{equation}

Accordingly, the final gradients of the loss function with respect to \(b_{1,k}\) and \(w_{1,k,\boldsymbol{l}}\) are:\hfill\mbox{}
\begin{equation}
\label{eq50}
\begin{gathered}
\frac{\partial L}{\partial b_{1,k}}
= \frac{1}{\sqrt{2\pi}}
\left( \sum_{s=0}^{d_{\mathrm{model}}-1} \delta_{n,s} \cdot w_{2,s,k} \right)
\cdot \left( \int_{-\infty}^{c_{1,k}} \exp\left(-\frac{x^2}{2}\right) dx + c_{1,k} \exp\left(-\frac{c_{1,k}^2}{2}\right) \right)\\[12pt]
\frac{\partial L}{\partial w_{1,k,\boldsymbol{l}}}
= \frac{1}{\sqrt{2\pi}}
\left( \sum_{s=0}^{d_{\mathrm{model}}-1} \delta_{n,s} \cdot w_{2,s,k} \right)
\cdot \left( \int_{-\infty}^{c_{1,k}} \exp\left(-\frac{x^2}{2}\right) dx + c_{1,k} \exp\left(-\frac{c_{1,k}^2}{2}\right) \right)
\cdot p_{\boldsymbol{l}}
\end{gathered}
\end{equation}

The corresponding parameter update formulas are:
\begin{equation}
\label{eq51}
b_{1,k}^{(t+1)} \leftarrow b_{1,k}^{(t)} - \eta_1 \frac{\partial L}{\partial b_{1,k}},\quad
w_{1,k,\boldsymbol{l}}^{(t+1)} \leftarrow w_{1,k,\boldsymbol{l}}^{(t)} - \eta_1 \frac{\partial L}{\partial w_{1,k,\boldsymbol{l}}}
\end{equation}

where \(\eta_1\) denotes the learning rate of the first classical linear layer.

\subsubsection{Parameter Updates for Quantum Layer}

Finally, the gradients of the loss function with respect to the parameters \(b_i\), \(w_{ij}\) and \(\theta_i\) of the quantum layer are computed, followed by the update of the parameters. The three parameter sets of the quantum layer are defined as follows:
\begin{equation}
\label{eq52}
\begin{gathered}
\boldsymbol{b} = \{b_i \mid i \in \{0,\;1,\;\dots,\;N-1\}\}\\
\boldsymbol{w} = \{w_{ij} \mid i,j \in \{0,\;1,\;\dots,\;N-1\},\; i<j\}\\
\boldsymbol{\theta} = \{\theta_i \mid i \in \{0,\;1,\;\dots,\;N-1\}\}
\end{gathered}
\end{equation}

Accordingly, the unitary evolution operators \(U_H\) and \(U_R\) defined above are combined into the total unitary evolution operator \(U\), i.e.:
\begin{equation}
\label{eq53}
U = U(\boldsymbol{b},\boldsymbol{w};\boldsymbol{\theta}) = U_R(\boldsymbol{\theta}) U_H(\boldsymbol{b},\boldsymbol{w})
\end{equation}

At this point, the final state \(|\psi_2\rangle\) modulated by the quantum layer can be briefly written as:\hfill\mbox{}
\begin{equation}
\label{eq54}
|\psi_2\rangle = U |\psi_0\rangle = U_R U_H |\psi_0\rangle
\end{equation}

Equations~\ref{eq53} and~\ref{eq54} show that the entire quantum circuit can be combined into a total unitary evolution operator \(U\), where \(\boldsymbol{b}\), \(\boldsymbol{w}\) and \(\boldsymbol{\theta}\) are the angular parameter sets of \(U\). Any change of an arbitrary parameter in the above three sets will cause the final state \(|\psi_2\rangle\) to change, thereby affecting the measurement results of all basis vectors. The measurement operator \(M_{\boldsymbol{l}}\) for the basis vector \(|\boldsymbol{l}\rangle\) is constructed as:
\begin{equation}
\label{eq55}
M_{\boldsymbol{l}} = |\boldsymbol{l}\rangle\langle\boldsymbol{l}|,\quad \sum_{\boldsymbol{l}=0}^{2^{N}-1} M_{\boldsymbol{l}} = \mathrm{I}
\end{equation}

After the final state \(|\psi_2\rangle\) is measured, the probability vector \(\boldsymbol{P}\) is obtained, where any element \(p_{\boldsymbol{l}}\) in this vector is the measurement expectation value corresponding to the measurement operator \(M_{\boldsymbol{l}}\):
\begin{equation}
\label{eq56}
p_{\boldsymbol{l}} = \langle M_{\boldsymbol{l}} \rangle = \langle \psi_2 | M_{\boldsymbol{l}} | \psi_2 \rangle \approx \frac{s t_{\boldsymbol{l}}}{s t_{\text{total}}}
\end{equation}

From the perspective of theoretical derivation, to obtain all elements in \(\boldsymbol{P}\), all \(2^N\) measurement operators \(M_{\boldsymbol{l}}\) are required to act on \(|\psi_2\rangle\) one by one for respective measurement and computation. In engineering practice, however, as shown in equation~\ref{eq56}, a total of \(st_{\text{total}}\) measurements can be performed at once. The count \(st_{\boldsymbol{l}}\) of outcomes corresponding to each basis vector \(|\boldsymbol{l}\rangle\) is recorded, and the measured frequency of this basis vector is further used to approximate the measurement probability.

The formulas for the gradients of the loss function with respect to the quantum layer parameters \(b_i\), \(w_{ij}\) and \(\theta_i\) are given as:
\begin{equation}
\label{eq57}
\begin{gathered}
\frac{\partial L}{\partial b_i} = \sum_{k=0}^{M-1}\left(\left(\sum_{s=0}^{d_{\text{model}}-1} \frac{\partial L}{\partial \hat{e}_{n,s}} \cdot \frac{\partial \hat{e}_{n,s}}{\partial c_{\text{GELU},k}}\right)\cdot \frac{\partial c_{\text{GELU},k}}{\partial c_{1,k}} \cdot \left(\sum_{\boldsymbol{l}=0}^{2^N-1}\frac{\partial c_{1,k}}{\partial p_{\boldsymbol{l}}}\cdot\frac{\partial p_{\boldsymbol{l}}}{\partial b_i}\right)\right)\\
\frac{\partial L}{\partial w_{ij}} = \sum_{k=0}^{M-1}\left(\left(\sum_{s=0}^{d_{\text{model}}-1} \frac{\partial L}{\partial \hat{e}_{n,s}} \cdot \frac{\partial \hat{e}_{n,s}}{\partial c_{\text{GELU},k}}\right)\cdot \frac{\partial c_{\text{GELU},k}}{\partial c_{1,k}} \cdot \left(\sum_{\boldsymbol{l}=0}^{2^N-1}\frac{\partial c_{1,k}}{\partial p_{\boldsymbol{l}}}\cdot\frac{\partial p_{\boldsymbol{l}}}{\partial w_{ij}}\right)\right)\\
\frac{\partial L}{\partial \theta_i} = \sum_{k=0}^{M-1}\left(\left(\sum_{s=0}^{d_{\text{model}}-1} \frac{\partial L}{\partial \hat{e}_{n,s}} \cdot \frac{\partial \hat{e}_{n,s}}{\partial c_{\text{GELU},k}}\right)\cdot \frac{\partial c_{\text{GELU},k}}{\partial c_{1,k}} \cdot \left(\sum_{\boldsymbol{l}=0}^{2^N-1}\frac{\partial c_{1,k}}{\partial p_{\boldsymbol{l}}}\cdot\frac{\partial p_{\boldsymbol{l}}}{\partial \theta_i}\right)\right)
\end{gathered}
\end{equation}

The results for all terms in Equation~\ref{eq57} are given as:
\begin{equation}
\label{eq58}
\begin{gathered}
\sum_{s=0}^{d_{\text{model}}-1} \frac{\partial L}{\partial \hat{e}_{n,s}} \cdot \frac{\partial \hat{e}_{n,s}}{\partial c_{\text{GELU},k}} = \sum_{s=0}^{d_{\text{model}}-1} \delta_{n,s} \cdot w_{2,s,k}\\[4pt]
\frac{\partial c_{\text{GELU},k}}{\partial c_{1,k}} = \frac{1}{\sqrt{2\pi}}\left( \int_{-\infty}^{c_{1,k}} \exp\left(-\frac{x^2}{2}\right) dx + c_{1,k}\exp\left(-\frac{{c_{1,k}}^2}{2}\right) \right)\\[4pt]
\frac{\partial c_{1,k}}{\partial p_{\boldsymbol{l}}} = w_{1,k,\boldsymbol{l}}\\[4pt]
\frac{\partial p_{\boldsymbol{l}}}{\partial b_i} = \frac{\partial \langle M_{\boldsymbol{l}} \rangle}{\partial b_i} = \langle M_{\boldsymbol{l}} \rangle\left(b_i + \frac{\pi}{4}\right) - \langle M_{\boldsymbol{l}} \rangle\left(b_i - \frac{\pi}{4}\right) \approx \frac{st_{\boldsymbol{l}}^{b_i^+} - st_{\boldsymbol{l}}^{b_i^-}}{st_{\text{total}}}\\[4pt]
\frac{\partial p_{\boldsymbol{l}}}{\partial w_{ij}} = \frac{\partial \langle M_{\boldsymbol{l}} \rangle}{\partial w_{ij}} = \langle M_{\boldsymbol{l}} \rangle\left(w_{ij} + \frac{\pi}{4}\right) - \langle M_{\boldsymbol{l}} \rangle\left(w_{ij} - \frac{\pi}{4}\right) \approx \frac{st_{\boldsymbol{l}}^{w_{ij}^+} - st_{\boldsymbol{l}}^{w_{ij}^-}}{st_{\text{total}}}\\[4pt]
\frac{\partial p_{\boldsymbol{l}}}{\partial \theta_i} = \frac{\partial \langle M_{\boldsymbol{l}} \rangle}{\partial \theta_i} = \langle M_{\boldsymbol{l}} \rangle\left(\theta_i + \frac{\pi}{4}\right) - \langle M_{\boldsymbol{l}} \rangle\left(\theta_i - \frac{\pi}{4}\right) \approx \frac{st_{\boldsymbol{l}}^{\theta_i^+} - st_{\boldsymbol{l}}^{\theta_i^-}}{st_{\text{total}}}
\end{gathered}
\end{equation}

Equation~\ref{eq58} employs the parameter-shift rule (PSR)\cite{mitarai2018quantum} in the computation of partial derivatives for the quantum layer and provides formulas for approximating partial derivative values in experiments. The term with a positive sign in the numerator corresponds to the outcome counts corresponding to each basis vector \(|\boldsymbol{l}\rangle\) after shifting the parameter by \(+\pi/4\) and executing the circuit \(st_{\text{total}}\) times. Meanwhile, the term with a negative sign represents the outcome counts corresponding to each basis vector \(|\boldsymbol{l}\rangle\) after shifting the parameter by \(-\pi/4\) and executing the circuit \(st_{\text{total}}\) times. In engineering practice, once the parameters are adjusted, only \(st_{\text{total}}\) measurements need to be performed. Then the outcome counts corresponding to each basis vector \(|\boldsymbol{l}\rangle\) can be read, and the measured frequency of the basis vector is adopted to approximate its measurement probability.

The corresponding parameter update formulas are:
\begin{equation}
\label{eq59}
b_i^{(t+1)} \leftarrow b_i^{(t)} - \eta \frac{\partial L}{\partial b_i},\quad
w_{ij}^{(t+1)} \leftarrow w_{ij}^{(t)} - \eta \frac{\partial L}{\partial w_{ij}},\quad
\theta_i^{(t+1)} \leftarrow \theta_i^{(t)} - \eta \frac{\partial L}{\partial \theta_i}
\end{equation}

where \(\eta\) denotes the learning rate of the quantum layer.

\subsection{Generation Process}

The basic generation flow is as follows. Based on the quantum state encoding of the prompt, the trained model performs one forward propagation to obtain the \(d_{\text{model}}\)-dimensional predicted embedding for the next token. The token closest to this embedding is retrieved from the embedding matrix as the generated token. Subsequently, the newly generated token is appended to the sequence to predict the token after the next token. Text is produced via autoregressive generation until the termination condition is met.

Let the quantum state encoded by the prompt be \(|\psi_{0}^{\text{prompt}}\rangle\). After the execution of the trained quantum layer and, subsequently, the trained first and second classical linear layers, a \(d_{\text{model}}\)-dimensional predicted token embedding \(\hat{\boldsymbol{e}}_n\) is obtained. Next, the dot product between \(\hat{\boldsymbol{e}}_n\) and all token vectors in the embedding matrix is computed to yield the similarity score \(sim_t\):\hfill\mbox{}
\begin{equation}
\label{eq60}
sim_t = \hat{\boldsymbol{e}}_n \cdot \boldsymbol{e}_t,\quad t=0,1,\dots,V-1
\end{equation}

where the variable \(V\) denotes the number of tokens in the embedding matrix. Following the computation of the standard self-attention mechanism, the similarity scores are scaled to obtain the scaled similarity score \(sim^{\text{scaled}}_t\):
\begin{equation}
\label{eq61}
sim_t^{\mathrm{scaled}} = \frac{sim_t}{\sqrt{d_{\mathrm{model}}}},\quad t=0,1,\dots,V-1
\end{equation}

All tokens are sorted according to the scaled similarity scores. The top-\(m\) scaled similarity scores are selected to form the raw candidate similarity score set \(\boldsymbol{TP}^{\text{raw}}\):
\begin{equation}
\label{eq62}
\boldsymbol{TP}^{\mathrm{raw}} = \left\{ tp_1^{\mathrm{raw}},\, tp_2^{\mathrm{raw}},\, \dots,\, tp_m^{\mathrm{raw}} \,\bigg|\, tp_i^{\mathrm{raw}} \text{ is the }i\text{-th ranked } sim_t^{\mathrm{scaled}},\, m<V \right\}
\end{equation}

After removing tokens that fail to meet the filtering rules, the filtered candidate similarity score set \(\boldsymbol{TP}^{\text{filter}}\) of \(\mathit{sim}^{\text{scaled}}_t\) is obtained:
\begin{equation}
\label{eq63}
\boldsymbol{TP}^{\mathrm{filter}} = \left\{ tp_1^{\mathrm{filter}},\, tp_2^{\mathrm{filter}},\, \dots,\, tp_k^{\mathrm{filter}} \,\bigg|\, tp_j^{\mathrm{filter}} \text{ is the }j\text{-th ranked } sim_t^{\mathrm{scaled}} \text{ after filtering},\, k\le m \right\}
\end{equation}

After applying maximum shift to each element in \(\boldsymbol{TP}^{\text{filter}}\), softmax normalization is conducted to obtain the sampling probability \(\mathit{prob}_j\) of each token:
\begin{equation}
\label{eq64}
prob_j = \frac{\exp\left(tp_j^{\mathrm{filter}} - \max\limits_{1\le j''\le k}\left(tp_{j''}^{\mathrm{filter}}\right)\right)}
{\sum\limits_{j'=1}^{k}\exp\left(tp_{j'}^{\mathrm{filter}} - \max\limits_{1\le j''\le k}\left(tp_{j''}^{\mathrm{filter}}\right)\right)}
\end{equation}

Thus, the discrete probability distribution over each element in \(\boldsymbol{TP}^{\text{filter}}\) is obtained as \(\{\mathit{prob}_1, \mathit{prob}_2, \dots, \mathit{prob}_k\}\). Sampling is performed on the corresponding tokens according to this discrete distribution to obtain the \((n+1)\)-th token generated in this round. The generated token is concatenated with the previous \(n-1\) tokens to form a new sequence of \(n\) tokens. The same generation procedure is executed to obtain the \((n+2)\)-th token. This autoregressive text generation process is repeated iteratively until the termination condition is met.

\subsection{Algorithm}

Based on the above derivations, the training algorithm and generation algorithm of FC-QBM-LM are presented separately, as shown below.
\clearpage
\begin{figure}[!ht]
\centering
\label{alg1}
\includegraphics[width=1\textwidth]{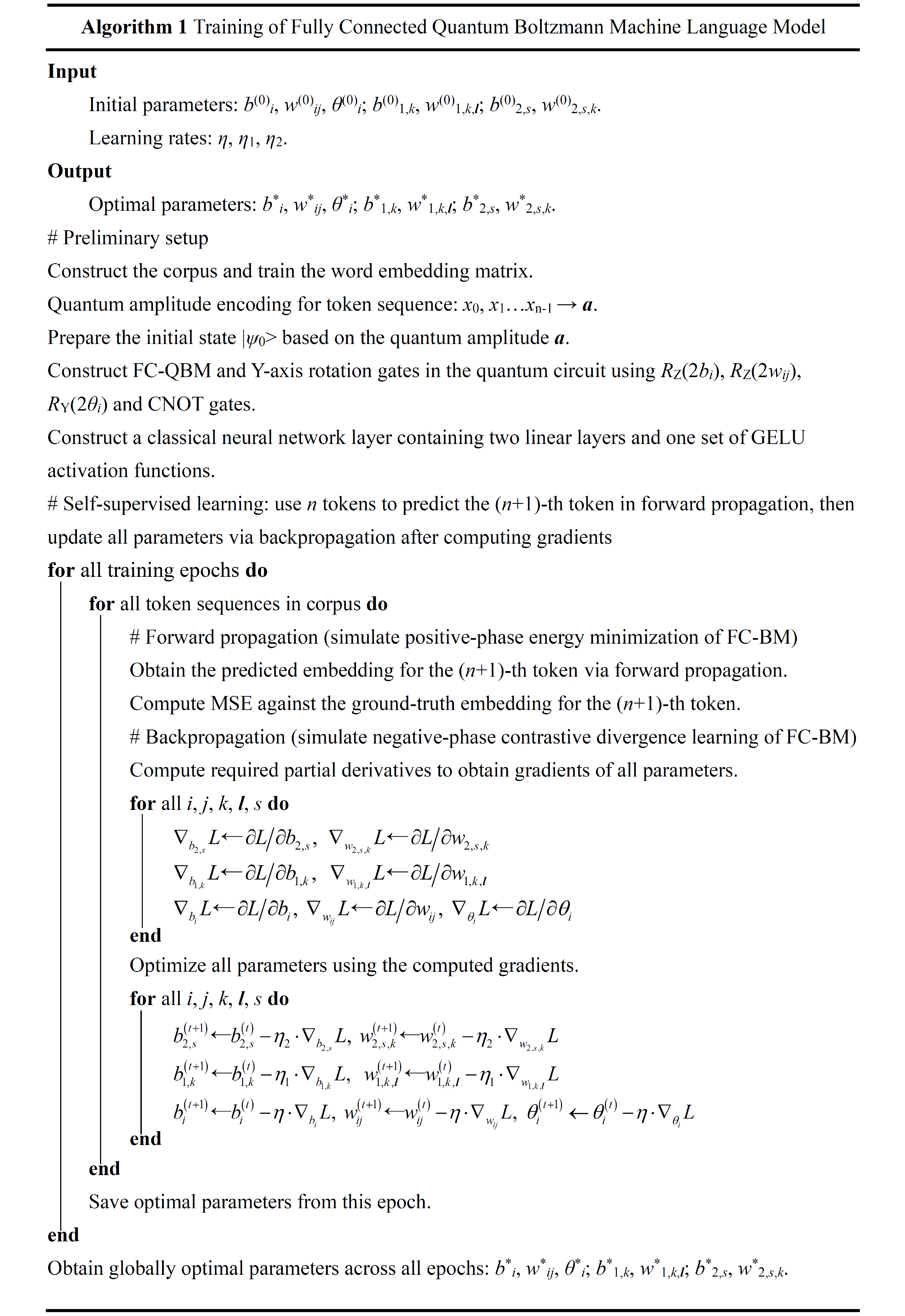}
\end{figure}

\clearpage
\begin{figure}[!ht]
\centering
\label{alg2}
\includegraphics[width=1\textwidth]{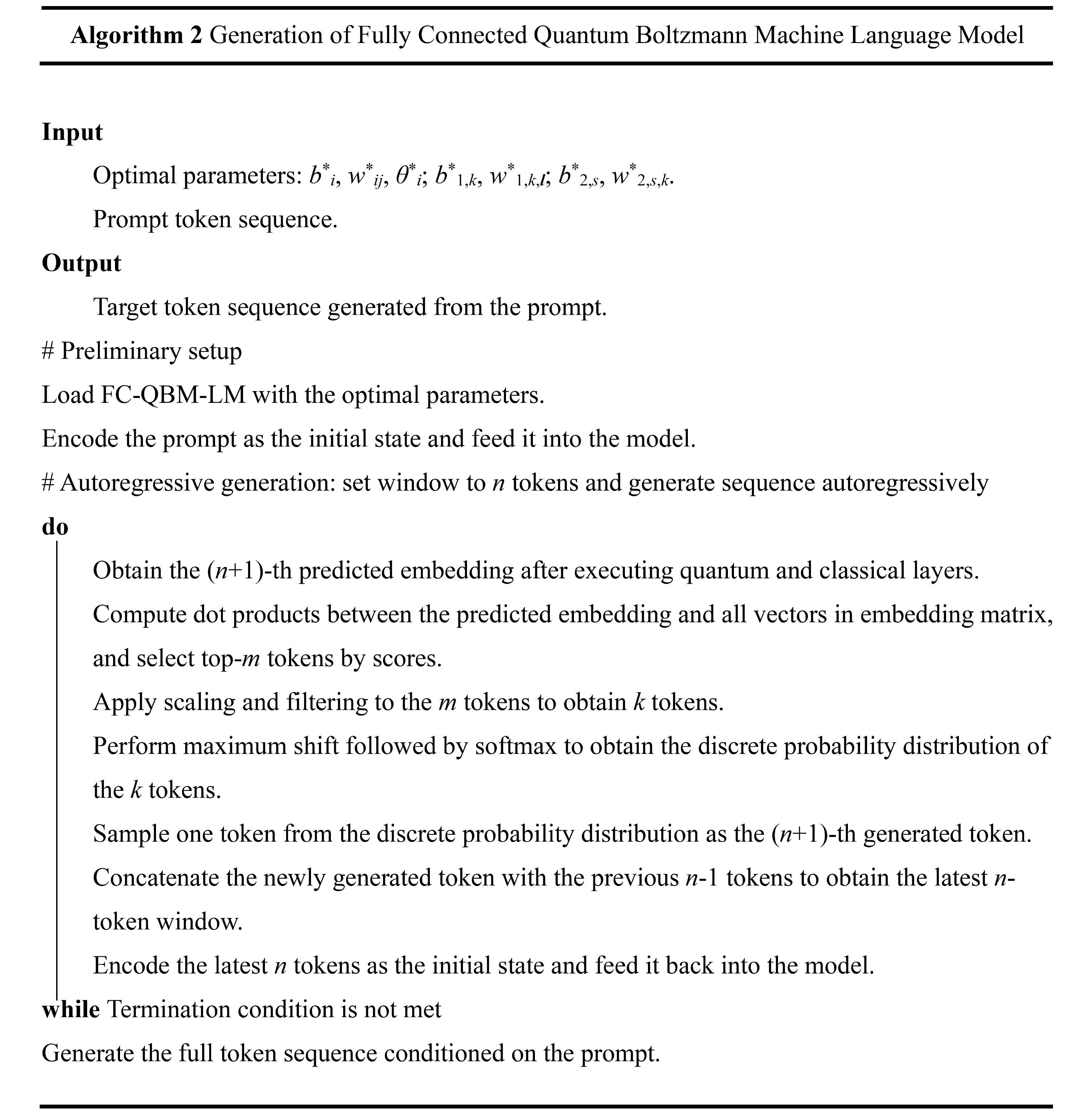}
\end{figure}
\vspace{1em}

\section{8-Qubit Fully Connected Quantum Boltzmann Machine Language Model}

In this work, \textit{Three Hundred Tang Poems} is adopted as the raw corpus for an 8-qubit FC-QBM-LM. Its text is cleaned and preprocessed to yield the training corpus. A word2vec model based on skip-gram is used to learn word embeddings, and an embedding matrix specific to Tang poetry with vocabulary size \(V=2481\) and embedding dimension \(d_\text{model}=25\) is constructed. Following the principle of traceability, all training code, execution logs and the final \texttt{.csv} file of the embedding matrix are properly preserved. 

The model consists of the quantum module and the classical module, and focuses on training and generation. Specifically, the quantum module consists of one quantum layer, and the classical module consists of two linear layers with a GELU activation function inserted in between. The number of tokens processed by the model at each time step is set to \(n=10\), i.e., the input token sequence is \(x_0, x_1, \dots, x_9\). Accordingly, an initial concatenated token vector \(\boldsymbol{TK}_0\) with dimension \(n\times d_\text{model}=250\) is obtained. Based on the above estimation method, the ceiling function is adopted to compute the number of qubits required in the quantum layer for encoding the 250-dimensional concatenated token vector:
\begin{equation}
\label{eq65}
N = \left\lceil \log_2\bigl(n \times d_{\text{model}}\bigr) \right\rceil
= \left\lceil \log_2 250 \right\rceil = 8
\end{equation}

Since the total number of basis vectors is \(2^8=256\), the amplitudes of the extra 6 dimensions are padded with zeros. Through the foregoing processing ‌procedure‌s, the 250-dimensional concatenated token vector \(\boldsymbol{TK}_0\) is converted into a 256-dimensional amplitude encoding vector \(\boldsymbol{a}\). Subsequently, \(\boldsymbol{a}\) is encoded into the 8-qubit quantum circuit to prepare the initial state \(|\psi_0\rangle\):
\begin{equation}
\label{eq66}
|\psi_0\rangle = \sum_{\boldsymbol{r}=00000000}^{11111111} a_{\boldsymbol{r}} |\boldsymbol{r}\rangle
\end{equation}

Next, FC-QBM and Y-axis rotation gates are constructed within the quantum circuit following the computation and derivation procedures presented earlier. The Hamiltonian expression corresponding to FC-QBM is:
\begin{equation}
\label{eq67}
H=\sum_{i=0}^{7} b_i \sigma_{\mathrm{Z}}^i + \sum_{i=0}^{7}\sum_{j=i+1}^{7} w_{ij} \sigma_{\mathrm{Z}}^i \sigma_{\mathrm{Z}}^j
\end{equation}

\begin{figure}[htbp]
\centering
\includegraphics[width=1.25\textwidth, angle=90]{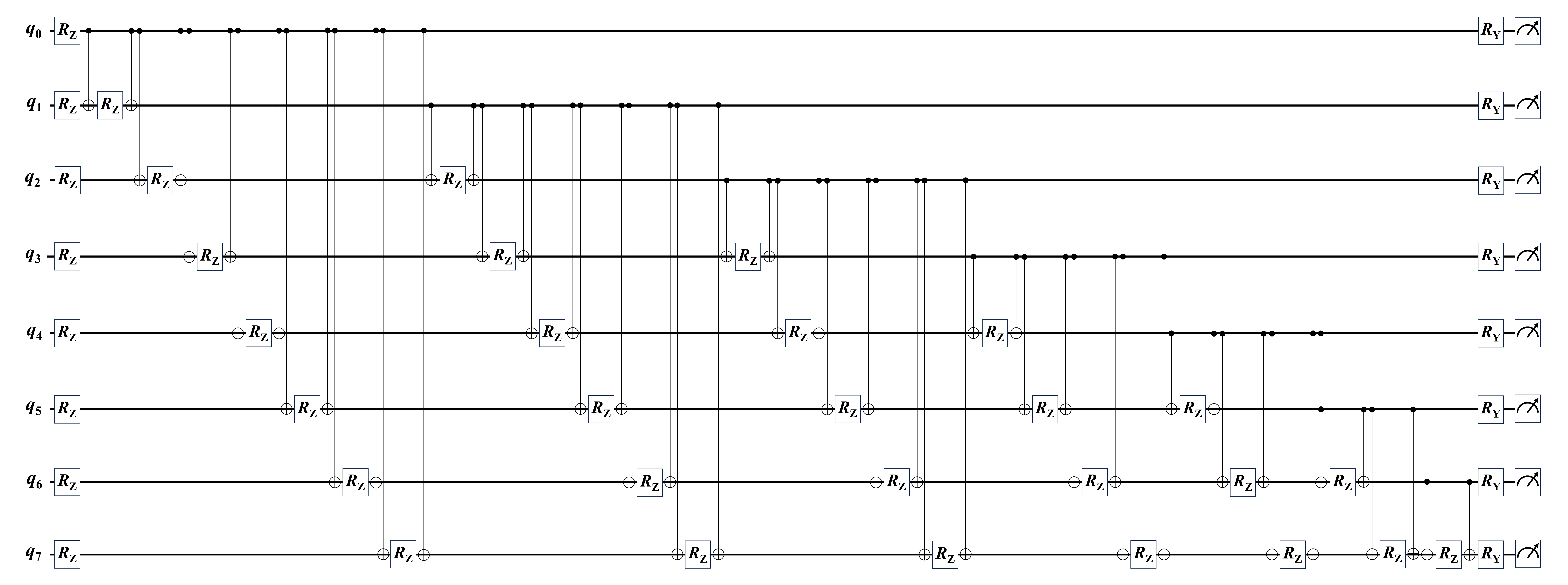}
\caption{Quantum Circuit of 8-Qubit Fully Connected Quantum Boltzmann Machine Language Model}
\label{fig2}
\parbox{\linewidth}{%
\raggedright
\setlength{\parindent}{0pt}\small
Trainable variational angles are assigned as follows: bias $R_{\mathrm{Z}}$ is parameterized by $2b_i$, entangling $R_{\mathrm{Z}}$ by weights $2w_{ij}$ and $R_{\mathrm{Y}}$ by $2\theta_i$.
}
\end{figure}

The structure of the quantum circuit is illustrated in Figure~\ref{fig2}. Its function is to implicitly implement positional encoding and self-attention computation through FC-QBM. FC-QBM is realized by \(R_\mathrm{Z}(2b_i)\), \(R_\mathrm{Z}(2w_{ij})\) and CNOT gates, while Y-axis rotation gates are implemented by \(R_\mathrm{Y}(2\theta_i)\) gates. The number of parameters \(w_{ij}\) is 28, and the numbers of \(b_i\) and \(\theta_i\) are each 8, leading to a total of 44 parameters in the quantum circuit. It should be noted that all angle parameters of rotation gates are omitted in Figure~\ref{fig2} due to space constraints.

After the quantum circuit is executed, the final state \(|\psi_2\rangle\) is obtained. A 256-dimensional probability vector \(\boldsymbol{P}=[p_{00000000}\ p_{00000001}\ \dots\ p_{11111111}]\) is further extracted from measurement outcomes. This probability vector is fed into the classical module to produce the next token embedding predicted by the model. The first classical linear layer fuses the features of \(\boldsymbol{P}\) and reduces its dimensionality, compressing it into a 128-dimensional vector:\hfill\mbox{}
\begin{equation}
\label{eq68}
\begin{gathered}
c_{1,k} = \sum_{\boldsymbol{l}=000000000}^{\boldsymbol{l}=111111111} w_{1,k,\boldsymbol{l}} p_{\boldsymbol{l}} + b_{1,k} \\
\boldsymbol{C}_1 = \big[\, c_{1,0}\;\; c_{1,1}\;\; \dots \;\; c_{1,127} \,\big]
\end{gathered}
\end{equation}

Here, the number of parameters \(b_{1,k}\) is 128, and the number of \(w_{1,k,\boldsymbol{l}}\) is \(128\times256=32768\). Thus, the total number of parameters in the first classical linear layer is 32896. Subsequently, elements in vector \(\boldsymbol{C}_1\) are nonlinearly transformed by the GELU activation function to yield the 128-dimensional vector \(\boldsymbol{C}_\mathrm{GELU}\):
\begin{equation}
\label{eq69}
\begin{gathered}
c_{\mathrm{GELU},k} = c_{1,k} \cdot \Phi\left(c_{1,k}\right) \\
\boldsymbol{C}_{\mathrm{GELU}} = \big[\, c_{\mathrm{GELU},0}\;\; c_{\mathrm{GELU},1}\;\; \dots \;\; c_{\mathrm{GELU},127} \,\big]
\end{gathered}
\end{equation}

Finally, the vector \(\boldsymbol{C}_\mathrm{GELU}\) is fed into the second classical linear layer to produce the 25-dimensional predicted token embedding \(\hat{\boldsymbol{e}}_{10}\):
\begin{equation}
\label{eq70}
\begin{gathered}
c_{2,s} = \sum_{k=0}^{127} w_{2,s,k} c_{\mathrm{GELU},k} + b_{2,s} \\
\hat{\boldsymbol{e}}_{10} = \big[\, \hat{e}_{10,0}\;\; \hat{e}_{10,1}\;\; \dots \;\; \hat{e}_{10,24} \,\big] = \big[\, c_{2,0}\;\; c_{2,1}\;\; \dots \;\; c_{2,24} \,\big]
\end{gathered}
\end{equation}

Here, the number of parameters \(b_{2,s}\) is 25, and the number of \(w_{2,s,k}\) is \(25\times128=3200\). Thus, the total number of parameters in the second classical linear layer is 3225.

As derived above, the total number of trainable parameters in the proposed model is 36165. Among them, the quantum circuit layer contributes only 44 parameters, while the two classical linear layers account for the remaining \(32896 + 3225=36121\) parameters. It is worth noting that the quantum layer uses only 44 parameters to implement positional encoding and self-attention computation, demonstrating that the quantum language model based on FC-QBM can substantially reduce parameter overhead.
\clearpage
\begin{table}[htbp]
  \centering
  \setlength{\tabcolsep}{23pt}
  \renewcommand{\arraystretch}{1}
  \caption{Parameter List of 8-Qubit Fully Connected Quantum Boltzmann Machine Language Model}
  \label{tab1}
  \begin{tabular}{lccc}
    \noalign{\hrule height 0.8pt}
    Layer & Parameter & Quantity & Subtotal \\
    \hline
    & $b_i$ & 8 & \\
    Quantum self-attention layer & $w_{ij}$ & 28 & 44 \\
    \phantom{Quantum self-attention layer} & $\theta_i$ & 8 & \\
    \raisebox{-1.2ex}{First classical linear layer} & $b_{1,k}$ & 128 & \\
    \phantom{First classical linear layer} & $w_{1,k,\boldsymbol{l}}$ & 32768 & \raisebox{2ex}{32896} \\
    \raisebox{-1.2ex}{Second classical linear layer} & $b_{2,s}$ & 25 & \\
    \phantom{Second classical linear layer} & $w_{2,s,k}$ & 3200 & \raisebox{2ex}{3225} \\
    Total & & 36165 & 36165 \\
    \noalign{\hrule height 0.8pt}
  \end{tabular}
\end{table}

During training, 10 tokens are sequentially read from the corpus of \textit{Three Hundred Tang Poems}. These tokens are encoded following the aforementioned algorithm and fed into the model. The forward propagation of the model is used to simulate the positive-phase energy minimization of FC-BM. Leveraging the properties of quantum computation, the quantum layer implicitly implements self-attention processing without matrix operations and outputs a probability vector encoding token positions and interactions. The classical layers further analyze this probability vector and predict the next plausible token embedding. In forward propagation, with all parameters initialized to random values, the model runs from the quantum layer to the subsequent classical layers and finally outputs the next predicted token embedding. Subsequently, parameter gradients are obtained via backpropagation, and all parameters are updated, simulating the negative-phase contrastive divergence learning process of FC-BM. The above process runs iteratively until the entire \textit{Three Hundred Tang Poems} corpus is traversed, marking the completion of one full training epoch.

In the generation stage, the prompt is first encoded into a quantum state and fed into the model. The trained quantum and classical layers perform one forward propagation to obtain a 25-dimensional predicted embedding for the next token. On this basis, the token with the smallest distance to this predicted embedding is retrieved from the embedding matrix trained on \textit{Three Hundred Tang Poems} and taken as the generated token. The newly generated token is then appended to the sequence to predict the subsequent token. The above generation process is repeated to produce text in an autoregressive manner until the termination condition is met.
\begin{figure}[!ht]
  \centering
  \makebox[\linewidth][l]{\hspace*{-70pt}\includegraphics[width=1.30\linewidth]{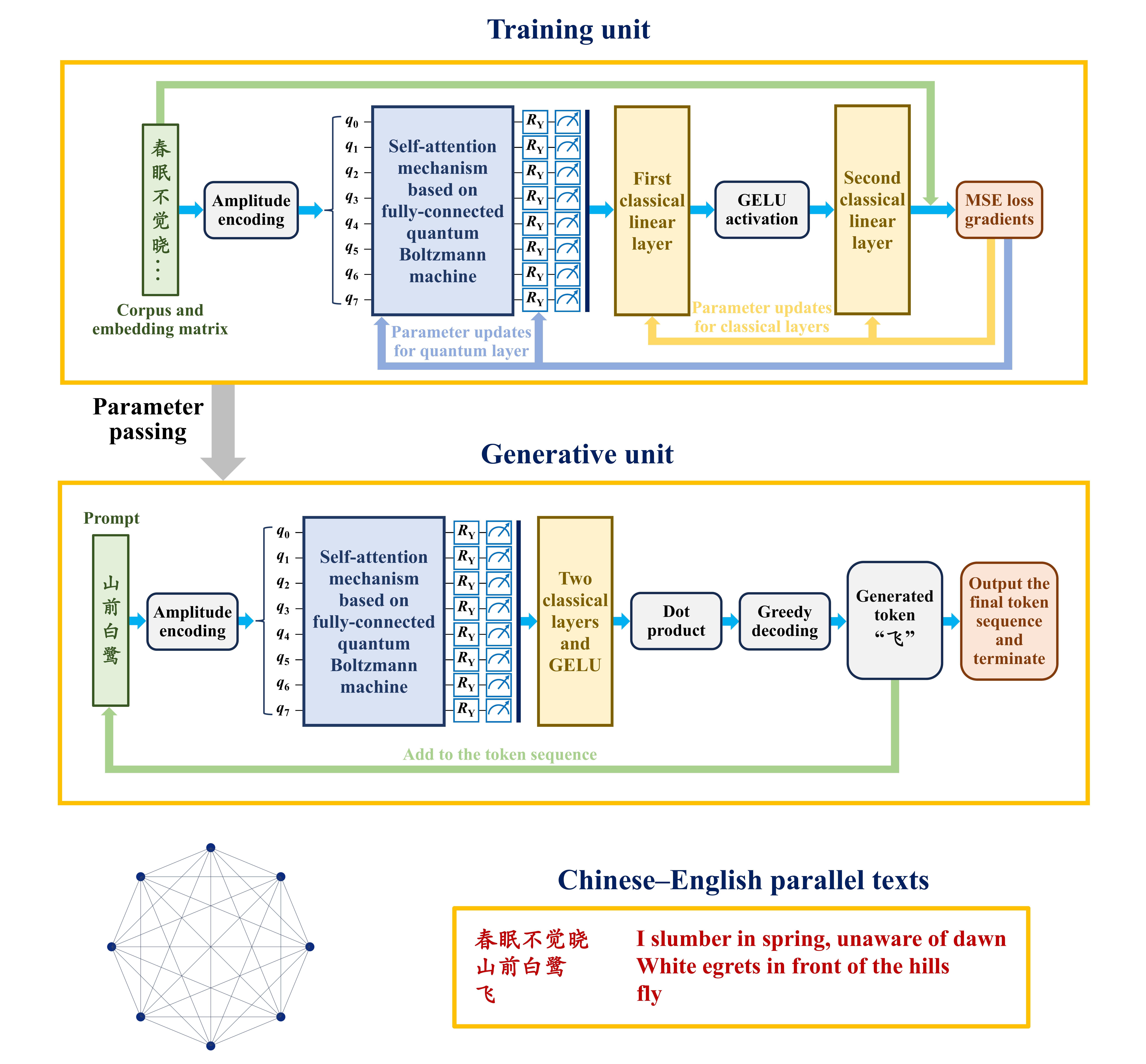}}
  \caption{System Block Diagram of 8-Qubit Fully Connected Quantum Boltzmann Machine Language Model}
  \label{fig3}
\end{figure}

\section{Experimental Results and Analysis}

To comprehensively evaluate the model performance, three sets of experiments are designed in this work, namely model convergence experiments, noise robustness experiments and five-character quatrain generation experiments. All experiments are implemented based on the PennyLane quantum simulator and the PyTorch machine learning framework. Following the principle of traceability, all source codes, execution logs, experimental data and model checkpoint files in \texttt{.pth} format are properly preserved.

\subsection{Model Convergence Experiments}

This set of experiments investigates the convergence performance of the model under noiseless conditions. Preprocessing of the original text of \textit{Three Hundred Tang Poems} yields a corpus of 23520 token samples, as well as a \(2481\times25\) embedding matrix corresponding to a vocabulary of 2481 distinct tokens. Checkpoints are employed in the training process. After each epoch, the model with the latest parameters is saved, and training for the next epoch resumes from this checkpoint. In addition, to conduct a thorough and accurate analysis of the model convergence, a control group with X-axis rotation gates at the circuit terminal is set up for comparison against the experimental group of the scheme proposed in this work, which adopts Y-axis rotation gates at the same terminal. Except for the rotation gates at the terminal, the two quantum circuits are identical in all other parts.

\begin{figure}[!ht]
  \centering
  \includegraphics[width=0.95\linewidth]{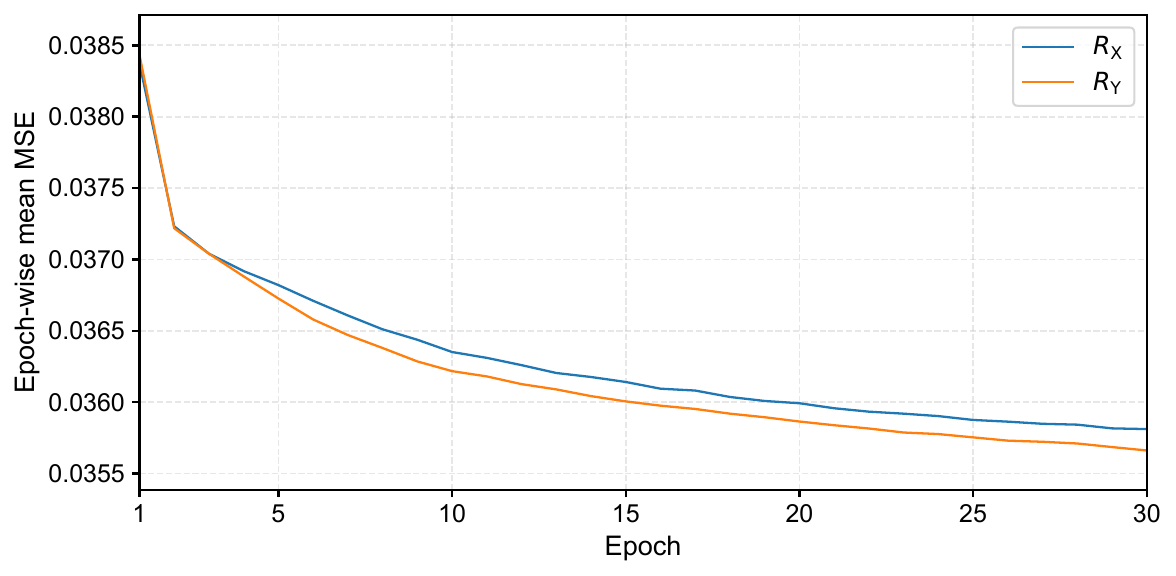} 
  \caption{Computational Results of Model Convergence Experiments}
  \label{fig4}
\end{figure}

The experiments use the \texttt{default.qubit} state-vector simulator built into PennyLane to simulate the quantum layer, and all computations are performed on a ThinkPad T14p Gen 2 laptop. A single epoch takes approximately 4.5 hours on average, and the model is trained for 30 epochs in total, yielding a total training time of \(2\times30\times4.5 = 270\) hours across all epochs for both groups. To quantitatively evaluate the convergence effect of the model, the arithmetic mean of the MSE over all token samples in each epoch is taken as the performance metric for that epoch, and the ‌corresponding curves are shown in Figure~\ref{fig4}.
\clearpage

From the curves in Figure~\ref{fig4}, it can be observed that under noiseless conditions: (1) As training epochs increase, the epoch-wise mean MSEs of both groups decrease progressively, with the curves becoming gradually less steep, demonstrating the favorable convergence performance of both groups. (2) In comparison, the quantum circuit with \(R_\mathrm{Y}\) gates at the terminal proposed in this work achieves noticeably better convergence performance than that terminated with \(R_\mathrm{X}\) gates.

\subsection{Noise Robustness Experiments}

In the current noisy intermediate-scale quantum (NISQ) era, noiseless quantum computing devices are yet to emerge. Noise exerts a significant impact on the precision of quantum computing. Therefore, this set of experiments specifically investigates the robustness of the proposed model against noise perturbations.

The baseline noise level is defined as 0.5\% for single-qubit gates and 1.0\% for two-qubit gates. This definition takes into account the global landscape of commercial superconducting quantum hardware as of August 2026, upon the completion of all experiments, and matches the real-world scenario where early-generation, mid-generation and state-of-the-art devices run in parallel. Compared with the noise setting of 0.5\% for single-qubit gates and 2.0\% for two-qubit gates adopted in a prior study\cite{liu2026breaking}, this work reduces the baseline noise for two-qubit gates by one percentage point after the latest comprehensive hardware performance evaluation.

To accurately reconstruct real-world noise scenarios, this work adopts the \texttt{default.mixed} mixed-state qubit simulator in PennyLane and simulates real hardware noise based on depolarizing channels. It should be noted that this scheme relies on density matrix evolution, whose computational complexity and resource consumption are substantially higher than those of conventional state-vector simulation. Under the 8-qubit configuration in this work, the data size of the density matrix can theoretically reach \(2^8=256\) times that of state-vector simulation, and matrix evolution together with noise simulation introduces additional computational overhead. This inference is also validated by preliminary tests prior to the formal experiments. For this reason, only the first eight poems in \textit{Three Hundred Tang Poems}, containing a total of 501 token samples, are used for computations in the noise robustness experiments.

\begin{figure}[htbp]
\centering
\vspace{40pt}
\makebox[\linewidth][l]{\hspace*{-45pt}\includegraphics[width=1.2\linewidth]{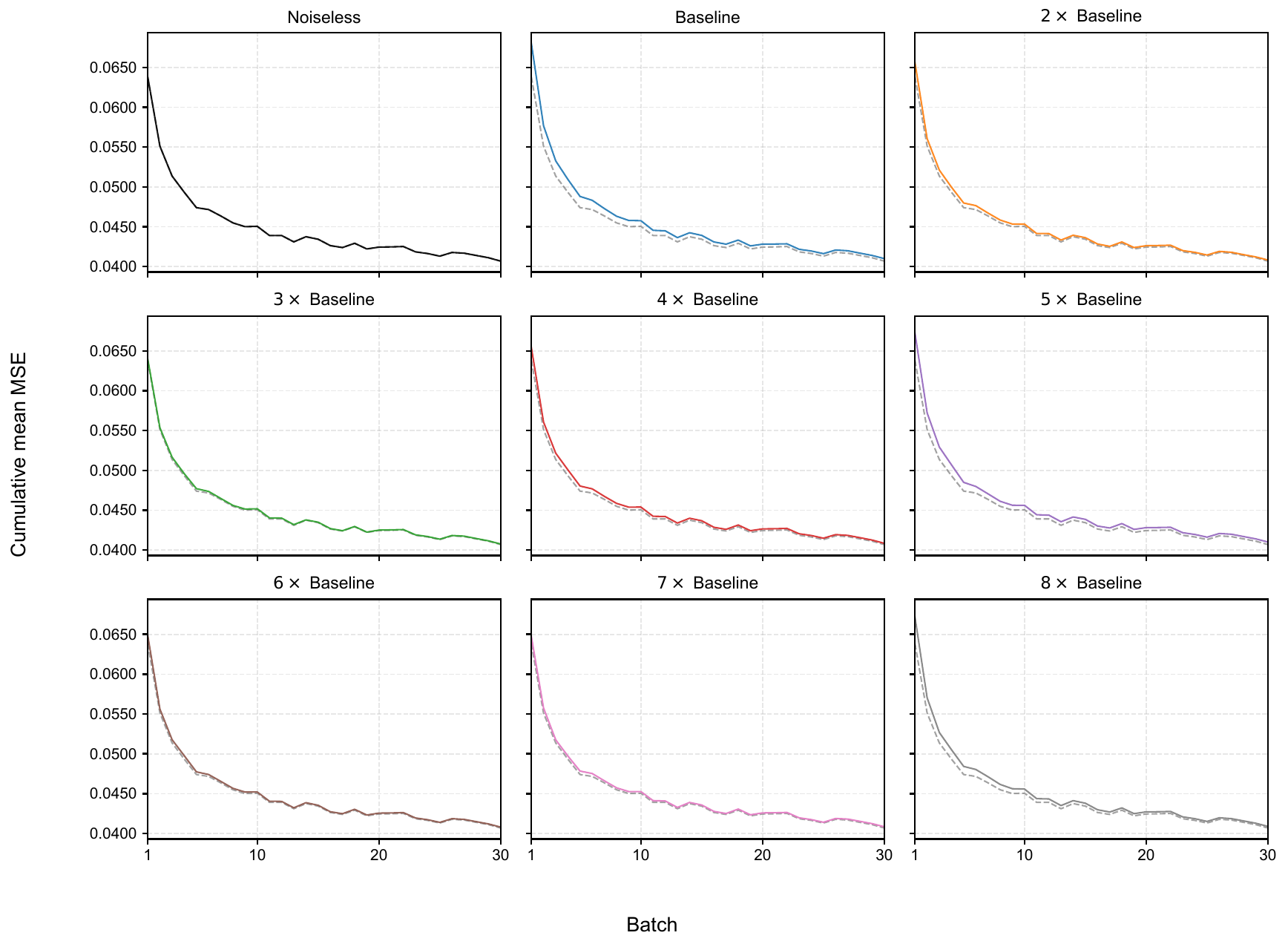}}
\caption{Computational Results of Noise Robustness Experiments}
\label{fig5}
\parbox{\linewidth}{%
\raggedright
\setlength{\parindent}{0pt}\small
The experimental scenarios include the noiseless case, baseline noise, 2× baseline noise, 3× baseline noise, \(4\times\)baseline noise, \(5\times\)baseline noise, \(6\times\)baseline noise, \(7\times\)baseline noise and \(8\times\)baseline noise. Starting from the second subplot, solid lines denote the cumulative mean MSE over batches under different noise scenarios, while the dashed line represents that for the noiseless case.
}
\end{figure}

On this basis, this work further divides the 501 token samples into 30 batches. The first 29 batches each contain 17 tokens, and the last batch contains 8 tokens. Meanwhile, to deeply investigate the model robustness under different noise scenarios, this work expands the baseline noise setting of 0.5\% for single-qubit gates and 1.0\% for two-qubit gates to cover the noiseless case, \(2\times\)baseline noise, …, up to \(8\times\)baseline noise, yielding a total of nine experimental scenarios. All computational experiments are performed on the ThinkPad T14p Gen 2 laptop, and the final results are shown in Figure~\ref{fig5} and Table~\ref{tab2}. The computation for the noiseless scenario takes approximately 1.5 hours. For the remaining eight noisy scenarios, the average computation time per scenario is about 5.5 hours. Therefore, the total computational time for this set of experiments is approximately \(1.5 + 8\times 5.5 = 45.5\) hours.
\clearpage

\clearpage
\begin{table}[!ht]
	\centering
  \caption{Cumulative Mean of Mean Squared Error and Convergence Performance Retention Rate Under Different Noise Scenarios.}
  \label{tab2}
  \renewcommand{\arraystretch}{1.0}
  \footnotesize
  \setlength{\tabcolsep}{2pt}
	\makebox[\linewidth][1]{
\begin{tabular}{l@{\hspace{10pt}}c@{\hspace{17pt}}c@{\hspace{10pt}}ccccccc}
    \noalign{\hrule height 1pt}
     & Noiseless & Baseline & $2\times$Baseline & $3\times$Baseline & $4\times$Baseline & $5\times$Baseline & $6\times$Baseline & $7\times$Baseline & $8\times$Baseline \\
    \noalign{\hrule}
    Mean      & 0.040688 & 0.041006 & 0.040829 & 0.040756 & 0.040847 & 0.041006 & 0.040798 & 0.040848 & 0.040889 \\
    SD        & 0.009318 & 0.009547 & 0.009263 & 0.009146 & 0.009344 & 0.009490 & 0.009293 & 0.009183 & 0.009482 \\
    Retention & 1.000000 & 0.992245 & 0.996547 & 0.998332 & 0.996107 & 0.992245 & 0.997304 & 0.996083 & 0.995084 \\
    \noalign{\hrule height 1pt}
  \end{tabular}
}
  \par\raggedright
  \vspace{0.5em}
  \noindent Mean in the table denotes the mean of the cumulative mean MSE across all 30 batches, and SD represents the corresponding standard deviation. Both metrics are computed considering variations in token count within each batch. Retention refers to the convergence performance retention rate, defined as the ratio of Mean under the noiseless case to that under a noisy case, quantifying the preservation of the model's convergence performance under noise perturbations.
\end{table}

From Figure~\ref{fig5} and Tabale~\ref{tab2}, it can be concluded that under the small-sample regime, the model exhibits strong robustness against noise perturbations of varying intensities, maintaining over 99\% of the noiseless convergence performance. In particular, for all cases where the noise intensity exceeds the baseline, the model’s convergence performance is no worse than that for the baseline noise case. This experimental result is highly consistent with the findings of Kuzmin et al. (2024). They found that injecting moderate noise into quantum neural networks can function similarly to regularization in classical neural networks, substantially improving generalization ability, reducing prediction MSE and boosting convergence performance\cite{kuzmin2024method}. In addition, the classical backend layers designed in this work possess a powerful nonlinear mapping capability, enabling them to adaptively compensate for shifts induced by noise and further stabilize the convergence performance of the model. Admittedly, this experimental result might be a special phenomenon specific to the small-sample regime. However, due to the limited computing power of the laptop, only preliminary small-scale experiments can be conducted at present. Furthermore, more refined and comprehensive verification on real quantum devices will be carried out in future work when conditions permit.

\subsection{Five-Character Quatrain Generation Experiments}

This set of experiments examines the model’s generative capability through the task of generating five-character Tang quatrains. Five-character quatrains are classic metrical poetic forms that matured in the Tang Dynasty. A standard form has four lines of five characters each, totaling 20 characters, making them one of the most concise forms in classical Chinese poetry. Emerging from the Yuefu folk-song tradition of the Han and Wei Dynasties, this poetic form passed through centuries of systematic experimentation with tonal and metrical patterns during the Southern Dynasties, before attaining a fully defined set of formal conventions and aesthetic paradigms in the Tang Dynasty. Drawing on highly condensed imagery to evoke a profound, immersive artistic mood, it stands as a core representative category of short verse included in \textit{Three Hundred Tang Poems}, with a highly distinctive stylistic identity. For researchers in natural language processing, the five-character quatrain features both strict formal constraints and high semantic density, making it a highly valuable experimental benchmark for classical Chinese poetry generation tasks.

\begin{figure}[!ht]
  \centering
  \includegraphics[width=1\linewidth]{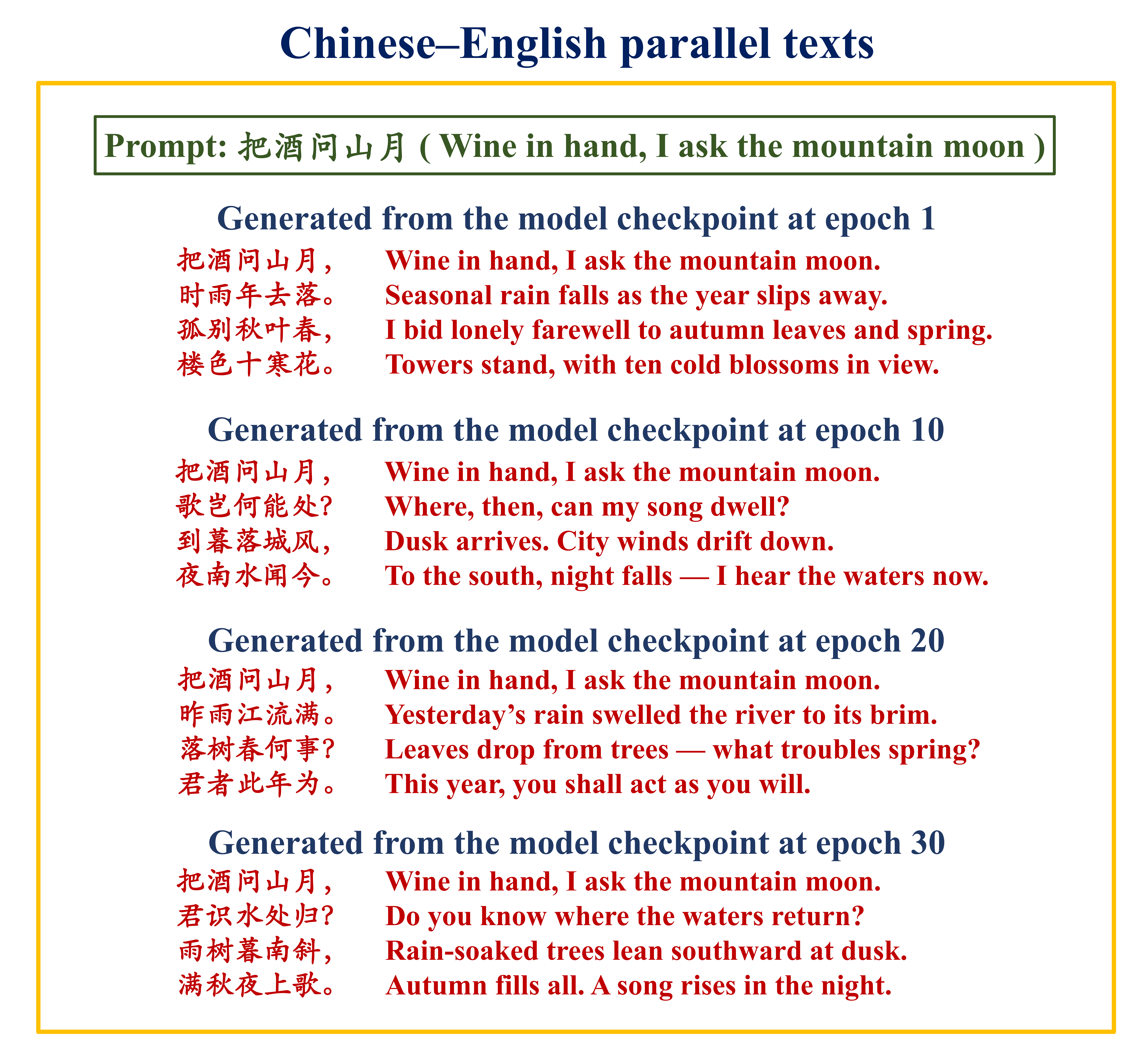} 
  \caption{Computational Results of Five-Character Quatrain Generation Experiments}
  \label{fig6}
\end{figure}

The prompt “Wine in hand, I ask the mountain moon” is specified as the opening line. Checkpoints at epoch 1, epoch 10, epoch 20 and epoch 30 from the first set of noiseless experiments are then used for the five-character quatrain generation tasks. The poetry generation code strictly follows the computational procedures described above. Any post-hoc manual modification, supplementation, polishing or other manipulations are strictly prohibited, so as to faithfully reflect the comprehensive performance of the model at different checkpoints. All computational experiments are conducted on the ThinkPad T14p Gen 2 laptop, and the experimental results are presented in Figure~\ref{fig6}.

\clearpage

From the experimental results shown in Figure~\ref{fig6}, the following conclusions can be drawn: (1) The model preliminarily exhibits associative and divergent capabilities, ‌indicating that‌ the self-attention mechanism based on FC-QBM functions effectively in the experiments. As observed from the results, the model can associate the prompt word “mountain moon” with related words such as “night”, “spring”, “autumn” and “river”. It can also map “ask” to interrogative words including “what” and “how”. Guided by the overall imagery of the full prompt “Wine in hand, I ask the mountain moon”, the model further forms associative and divergent links to relevant concepts such as homesickness, parting, passage of time and positive action. (2) The model possesses a certain capability to organize word order in accordance with human natural language conventions. It can be observed that the model is capable of roughly arranging words according to human natural language sequences and generating poems compliant with grammatical rules, demonstrating that positional encoding within the self-attention mechanism also takes effect in the model. (3) As training epochs increase, the coherence, fluency, antithesis and imagery rendering of the generated poems exhibit a trend of gradual improvement. (4) Constrained by architecture and parameter scale, the model constructed in this work is only a preliminary small-scale language model for verification, whose performance exhibits notable gaps when compared with professional large language models, leaving room for further exploration in subsequent research.

\section{Conclusion}

To address the limitations of the classical self-attention mechanism, a self-attention mechanism based on FC-QBM is designed in this work, and a quantum language model using the mechanism as the core engine is constructed. The findings are summarized as follows: (1) The encoding of token positions and the computation of token interactions are realized via FC-QBM, with \(R_\mathrm{Y}\) gates adopted for differentiated information extraction, and this design is mathematically equivalent to the classical self-attention mechanism. (2) Implicit computation for self-attention is realized by unitary evolution acting on semantic quantum states. Continuous semantic streams do not need to be decomposed into discrete matrices for explicit matrix multiplication, and quantum computing advantages can be leveraged to accelerate the overall process. Measurement results obtained after quantum circuit execution naturally satisfy the normalization condition without requiring softmax computation. Under the 8-qubit setting, the quantum layer implements the entire positional encoding and self-attention computation pipeline with only 44 parameters. (3) Under noiseless conditions, the model exhibits favorable convergence performance. Furthermore, the quantum circuit proposed in this work that terminates with \(R_\mathrm{Y}\) gates achieves better convergence performance than the counterpart terminated with \(R_\mathrm{X}\) gates. Under the small-sample regime and across eight distinct noise intensity levels, the model demonstrates strong noise robustness, retaining over 99\% of its noiseless convergence performance. (4) Experiments on five-character quatrain generation reveal that the model preliminarily exhibits three core capabilities: associative and divergent thinking, content generation across imagery and human-aligned word ordering. These results validate the effectiveness of the proposed self-attention mechanism based on FC-QBM within the model.

	\bibliographystyle{unsrtnat}
	\bibliography{references}
	
\end{document}